\documentclass{aa}
\pdfoutput=1
\usepackage[varg]{txfonts}
\newcommand{\kms}{\mbox{km\ s}$^{-1}$}

\newcommand{\msun}{M_\odot}

\newcommand{\msu}{M^{\rm su}}
\newcommand{\rsu}{R^{\rm su}}
\newcommand{\vsu}{V^{\rm su}}
\newcommand{\tsu}{T^{\rm su}}

\newcommand{\rs}{r_{\rm s}}

\newcommand{\Mvir}{M_{\rm vir}}

\newcommand{\Dvir}{\Delta_{\rm vir}}

\newcommand{\fc}{f_c}
\newcommand{\gc}{g_c}

\newcommand{\Om}{\Omega_{\rm m}}

\newcommand{\Omo}{\Omega_{{\rm m},0}}

\newcommand{\Olo}{\Omega_{\Lambda,0}}

\newcommand{\magal}{\textsc{MaGaLie}}
\newcommand{\nemo}{\textsc{NEMO}}

\begin{document}

%%%%%%%%%%%%%%%%%%%%%%%%%%%%%%%%%%%%%%%%%%%%%%%%%%

%\title[Major merger time-scales]{The time-scales of major mergers from
%  simulations of isolated binary galaxy collisions}

\title{The time-scales of major mergers from simulations of isolated
  binary galaxy collisions}

% The list of authors, and the short list which is used in the headers.
% If you need two or more lines of authors, add an extra line using \newauthor
\titlerunning{Major merger time-scales}
\authorrunning{Solanes, Perea, and Valent\'{i}-Rojas}
\author{J.~M.\ Solanes
  \inst{1}
  \and
  J.~D.\ Perea
  \inst{2}
  \and
  G.\ Valent\'{\i}-Rojas
  \inst{1}}

%\author[J.~M.\ Solanes,  J.~D.\ Perea and G.\ Valent\'{\i}--Rojas]{\parbox{\textwidth}{
%Jos\'e M.\ Solanes$^{1}$\thanks{E-mail: \texttt{jm.solanes@ub.edu}},
%Jaime D.\ Perea$^{2}$ and Gerard Valent\'{\i}--Rojas$^{1}$}\vspace{0.4cm}

\institute{Departament de F\'{\i}sica Qu\`antica i Astrof\'{\i}sica and
Institut de Ci\`encies del Cosmos (ICCUB), Universitat de
Barcelona. C.\ Mart\'{\i} i Franqu\`es, 1; E--08028~Barcelona,
Spain
%\email{jm.solanes@ub.edu}
\and
Instituto de Astrof\'{\i}sica de Andaluc\'{\i}a,
IAA--CSIC. Glorieta de la Astronom\'{\i}a, s/n; E--18008~Granada, Spain\\
%\email{jaime@iaa.es}
}

\offprints{J. M.\ Solanes, \email{jm.solanes@ub.edu}}

% These dates will be filled out by the publisher
%\date{To be submitted to A&A. Do not circulate.} % slugcomment
\date{Submitted to \aap}

% Relative path of figures folder
%\graphicspath {{figs/}}

\abstract{A six-dimensional parameter space based on high-resolution
  numerical simulations of isolated binary galaxy collisions has been
  constructed to investigate the dynamical friction time-scales,
  $\tau_{\rm mer}$, for major mergers. Our experiments follow the
  gravitational encounters between $\sim 600$ pairs of similarly
  massive late- and early-type galaxies with orbital parameters
  compliant to the predictions of the $\Lambda$CDM cosmology. We
  analyse the performance of different schemes for tracking the
  secular evolution of mergers, finding that the product of the
  intergalactic distance and velocity is best suited to identify the
  time of coalescence. In contrast, a widely used merger time
  estimator such as the exhaustion of the orbital spin is shown to
  systematically underpredict $\tau_ {\rm mer}$, resulting in relative
  errors that can reach $60\%$ for nearly radial
  encounters. Regarding the internal spins of the progenitors, we find
  that they can lead to total variations in the merger times larger
  than $30\%$ in highly circular encounters, whereas only that of the
  principal halo is capable of modulating the strength of the
  interaction prevailing throughout a merger. The comparison of our
  simulated merger times with predictions from different variants of a
  well-known fitting formula has revealed an only partially
  satisfactory agreement, which has led us to recalculate the values
  of the coefficients of these expressions to obtain relations that
  fit perfectly major mergers. The observed biases between data and
  predictions, that do not only apply to the present work, are
  inconsistent with expectations from differences in the degree of
  idealization of the collisions, their metric, spin-related biases, or
  the simulation set-up. This hints to a certain lack of accuracy of
  the dynamical friction modelling, arising perhaps from a still not
  quite complete identification of the parameters governing 
  orbital decay.}

\keywords{galaxies: haloes -- galaxies: interactions -- galaxies:
  kinematics and dynamics -- methods: numerical }
\label{firstpage}
\maketitle

\section{Introduction}\label{goals}

A consequence of the hierarchical nature of structure formation in a
cold dark matter (CDM) universe is that dark matter haloes are built
through the merging of earlier generations of less massive haloes. To
cause gravitationally bound interactions between extended,
self-gravitating haloes (and their galaxies) typically requires a
dissipative process that reduces the initial orbital energies of the
colliding objects. Dynamical friction fulfils this role by
transferring energy and momentum from their relative motion to
internal degrees of freedom. As a result, the orbits of interacting
galaxies evolve and may eventually decay and culminate in a merger.

The accurate determination of the merging time-scales driven by
dynamical friction is therefore a centrepiece of studies of the
evolution of galaxies. For example, in semi-analytic models of galaxy
evolution the merger times are directly involved in the assessment of
many fundamental quantities such as the luminosity/stellar mass
function, the amount of gas available to form new stars, the star
formation rate, the abundance of first-ranked galaxies, and the
distribution of galaxy sizes, metallicities, colours and morphologies
\citep[e.g.][]{KWG93,Lac01,RF10,Hir12,Ben12,Mun15}. Among all
galaxy-galaxy interactions, the class of so-called major mergers,
involving pairs of objects with similar mass, are specially relevant
as they play a central role, not only in the assembly history of
galaxies \citep*[e.g.][]{Lid13,Tas14}, but also in that of their
central supermassive black holes (SMBH) and the subsequent triggering
of nuclei activity \citep*[e.g.][]{FVD09,Yu11,Liu11,Kos12,Cap17}.

Virtually all formulas currently used to estimate the merger time,
$\tau_{\rm mer}$ are based on the idealized \citeauthor{Cha43}'s
(\citeyear{Cha43}) description of the deceleration caused by dynamical
friction on a point mass (representing the satellite) travelling
through an infinite, uniform and collisionless medium (representing
the host). They are mostly prescriptions inferred from the analytic or
semi-analytic modelling of mergers
\citep*[e.g.][]{LC93,vBos99,Taf03,Gan10,PGR15,Sil16} or parametric
equations tuned by direct comparison with the outcome of numerical
simulations of collisions of live galaxy pairs
\citep[e.g.][]{BKMQ08,Jia08,Jus11,McCa12,Vil13}, or both
\citep*[e.g.][]{CMG99}. However, although several decades of studies
of galactic mergers have allowed to reach a general consensus on what
should be the most relevant parameters in any process of this type,
the extend of their impact is still a matter of debate. Factors such
as continued mass losses due to tidal interactions, the drag force
exerted by tidal debris, the re-accretion of some of this material
onto the colliding galaxies, or the mutual tidal distortion of their
internal structures, to name just some, are elements that introduce
nontrivial uncertainties in any attempt to calculate $\tau_{\rm mer}$
using analytical expressions. To these difficulties motivated by
physical complexity one must add the lack of a one-hundred-per cent
standard methodology regarding the way mergers are tracked. The
starting point of this work deals precisely with the definition of a
new metric for calculating $\tau_{\rm mer}$ -- an obligatory first
step to compare the outcomes of different experiments under equal
conditions and derive universally valid formulas -- suitable for major
mergers. With this aim, we have run a suite of near 600
high-resolution $N$-body simulations of isolated major mergers, with
which we further explore the dependence of the merger time on a range
of orbital parameters, and progenitors' mass-ratios, spins and
morphologies, representative of such systems.

Several reasons justify the realization of this type of research with
controlled experiments. On one hand, because they involve a
substantial increase in resolution with respect to what would feasible
in an entirely cosmological context -- at least for samples of binary
collisions as large as ours --, allowing both host and satellites to
have more realistic internal structures. On the other hand, because we
aim to extend former studies focussing on the duration of mergers from
the standpoint of isolated systems by correcting their main
weaknesses. What we mean is, for instance, the systematic adoption of
ad-hoc initial conditions, such as strongly radial orbits and small
pericentric distances. Although there is some physical basis in this
approach, above all is the intention to lead to fast merging that
saves CPU time \citep[e.g.][]{KB06}. In addition, it is not uncommon
that the initial equilibrium galaxy models of controlled
investigations of dynamical friction, especially the oldest ones, do
not conform to the paradigm currently accepted for these systems by
including haloes with a parametrization of their internal structure
(e.g.\ isothermal) derived form first principles and, sometimes, with
the just minimum extension required by the observed flatness of the
rotation curves of galactic discs.

While today it is not difficult to find pre-prepared simulations of
binary collisions dealing with realistic galaxy models that satisfy
observational constraints \citep[e.g.][]{BKMQ08,VWa12,   Vil13,Bar16,Cap17}, the question of the assignment of initial
orbital conditions is an issue that is still solved guided more by
common wisdom rather than by specific evidence. Nevertheless, this
does not have to be the case. If the results of binary mergers served
at the time as a guide for the first fully cosmological experiments,
we can now use the latter as feedback to establish realistic initial
conditions for the former and thus to compare the results of both
types of simulations on a roughly equal footing. In this regard, we
wish to emphasize that the present work represents the first study of
dynamical friction based on gravitational encounters of well-resolved
pairs of galaxies resembling local objects whose set-up relies
entirely on initial conditions inferred from large-scale-structure
simulations carried out in the framework of the standard concordant
$\Lambda$CDM model, including those that incorporate the baryon
physics \citep[e.g.][]{KB06,Jia08,Bry13}.

Dealing with stand-alone systems also enables both to isolate the
effects of the different parameters that regulate dynamical friction
and to make cleaner and more accurate measurements than those that are
possible amidst a large-scale structure formation context. This is why
we compare our measured time-scales with predictions arising from
different variants of a physically-motivated approximating relation
widely used in both isolated and cosmological runs. Our original goal
was to establish whether, when the initial conditions governing
mergers are similar, so is their duration, regardless of the local
distribution of matter around the galaxies. What we have found,
however, has been a relatively important bias between measured and
estimated times that we had not anticipated, and whose magnitude and
behaviour surpass what would be expected from differences in the
degree of idealization of the merger context (isolated versus
cosmological), or in the metric, or in the configuration of the
simulations. The present investigation is completed by using our
simulations to recalculate the coefficients of the most popular
versions of the merger-time formula with the aim of providing
analytical prescriptions suitable for major mergers.

This then is the layout of the paper. In Section~\ref{simulations} we
present our suite of simulated binary collisions. After discussing
different alternatives for tracking major mergers and measuring their
length, the new metric for major mergers is defined in
Section~\ref{tau_merge}, while in Section~\ref{parameters} we explore
the dependence of $\tau_{\rm mer}$ on a series of parameters that are
known to regulate the effectiveness of dynamical friction. Finally, in
Section~\ref{tau_mer_scales}, we first compare the outcome of our
simulations with the predictions from a well-known set of parametric
models for $\tau_{\rm mer}$, and finish by readjusting the
coefficients of these models, so that they offer an improvement in the
predicted time-scales for major mergers. After outlining our
conclusions in Section~\ref{conclusionsI}, the present paper ends with
two Appendixes that run through details regarding the provision of
rotation to our haloes (Appendix~\ref{halorot}) and the strategy
followed to relate the initial conditions of our simulated galaxy
pairs with their subsequent orbital evolution driven by gravity
(Appendix~\ref{one-body eqs}).

Physical scales are calculated throughout the manuscript assuming a
standard flat $\Lambda$CDM cosmology with present-epoch values of the Hubble parameter, and total mass and dark energy density parameters equal to  
$H_0=70$~km~s$^{-1}$~Mpc$^{-1}$, $\Omo=0.3$ and $\Olo=0.7$, respectively.

\section{The simulations}\label{simulations}

In this section we explain the main ideas behind the numerical
methodology used to create our suite of galaxy mergers. Further
details regarding the galactic halo models used here can be found in
\citet{DS10}, \citet{Dar13}, and \citet{Sol16}.

\subsection{Models of galaxy pairs}\label{models}

Each galaxy pair consists of two non-overlapping extended spherical
haloes whose global properties (mass, spin, concentration, and density
profile of the dark component) are used to set the scalings of their
central baryonic (stellar) cores. Progenitor galaxies of late-type are
made of a spherical \citeauthor*{NFW97} (\citeyear{NFW97}; NFW) dark
matter halo, and a central exponential disc of stars surrounding a
non-rotating spherical \citet{Her90} stellar bulge. For the largest
progenitors, the bulge mass, $M_{\rm b}$, is taken equal to the
twenty-five per cent of the disc mass, $M_{\rm d}$, as they are
intended to represent a Hubble type Sb spiral, while $M_{\rm
  b}=0.1M_{\rm d}$ for the smallest ones, which imitate local Sc
galaxies \citep{Gra01}. Early-type galaxy models also consist of a
spherical NFW dark-matter halo and an inner non-rotating stellar
component represented by an oblate \citeauthor{Her90}'s profile with
an intrinsic flattening $c/a=0.7$, the most common value among local
elliptical galaxies \citep{Sta95}. In all cases, the mass of the
luminous component is taken equal to $5\%$ of the total galaxy
mass. Unlike other galaxy models that find it preferable to smoothly
taper off the density profiles using an exponential cut-off
\citep[e.g.][]{SW99,Bar16}, we have truncated sharply the stellar
distributions at a radius encompassing $95\%$ of the luminous mass and
the dark particle distributions at the halo virial radius,
redistributing the leftover mass within those radii. We expect our
calculations to be relatively insensitive to the details of the
truncation procedure.

As customary, we employ a numerically-convenient dimensionless system
of units in our simulations. We adopt $G=1$ for Newton's gravity
constant, with the virial mass, velocity and radius of the primary
galaxies set to one. Since these objects are assumed to have a total
mass equal to that commonly associated with a MW-sized galaxy in the
local universe, which is around $10^{12}\;h^{-1}\msun$ \citetext{see,
  e.g., \citealt*{KZS02,B-K09}, and references therein}, then,
according to the adopted cosmology, our models may be therefore
compared with real galaxies using the following physical scaling per simulation unit (su) for
mass, length, velocity and time: $\msu=1.42\times 10^{12}\msun$, $\rsu
= 297\;{\rm{kpc}}$, $\vsu=144~\mbox{km\ s}^{-1}$ and
$\tsu=\tau_{\rm dyn} (z=0)=2.02\;{\rm{Gyr}}$ (see eq.~(\ref{tau_dyn})), respectively. As a general rule, all quantities in this text are given in these 
units unless otherwise noted.

Parent galaxies of unit mass are modelled using a total of $210,000$
particles, while all our experiments adopt a fifty-fifty split in
number between luminous and dark bodies\footnote{This can be shown to
  provide a much better sampling of the former at the cost of a modest
  increase in the numerical heating that compensates the relatively
  moderate resolution of the experiments dictated by the large size of
  the sample (Section~\ref{simulations}).}. A single extra particle
representing a SMBH is added at the centre of each galaxy. The SMBH
particle masses are calculated assuming a mean ratio $\langle M_{\rm
  SMBH}/M_{\rm b}\rangle\sim 0.003$ typical of massive galaxies in the
local universe \citep[see, e.g.][and references therein]{Gra16}. We
set the Plummer equivalent softening length for the star particles to
30 pc, a value that in our simulations is comparable to the minimum
physical scale at which we can resolve the dynamics of the largest
remnants given the number of bodies adopted (see
Section~\ref{tau_merge}). In the case of the more massive bodies (DM
and SMBH), we take a softening length proportional to the square root
of their body mass to have the same maximum interparticle
gravitational force.

Our multicomponent galaxy models have been generated with the aid of a
much-improved version of the computationally efficient
\magal\ open-source code \citep{BKPG01} included in the \nemo\ Stellar
Dynamics
Toolbox\footnote{\texttt{http://bima.astro.umd.edu/nemo/}}. Our
upgraded version of the code allows simulated galaxies to begin very
close to equilibrium and require a negligible additional time to
settle. The new approximate solution adopted for the velocity field
causes a minor initial structural readjustment when the models are
evolved. In practice, this means that with a large enough number of
particles ($N\gtrsim 100,000$) transients damp out in less than one
half of a revolution at the half-mass radius, settling swiftly to
values close to those sought from the initial conditions. Once the
short transitional period is over, we have checked that the structural
and kinematic properties of all our galaxy models remain statistically
unchanged for as long as a Hubble time when they evolve in isolation,
even at the lowest resolution ($\sim 70,000$ particles per galaxy).

A neat rotation consistent with the adoption of a non-null value for
the dimensionless halo spin parameter, $\lambda$, is imparted to the
DM of our galaxies to incorporate the effect of large-scale
fluctuations. From Appendix~\ref{halorot} it can be seen that the net
fraction of dark matter halo particles that must spin in the
positive direction along the rotation axis (we choose this to be the
$Z$ axis) to produce a given $\lambda$ can be inferred from
\begin{equation}\label{fh}
\begin{split}
f_{\rm{h}}\equiv &
\frac{N_{\bullet,j^+_z}-N_{\bullet,j^-_z}}{N_\bullet}\approx \\ &
3\frac{\lambda}{\alpha\gc}\frac{[\ln(1+c)-c/(1+c)]^{5/2}}{[1-1/(1+c)^2-2\ln(1+c)/(1+c)]^{1/2}}\;,
\end{split}
\end{equation}
where $N_{\bullet}$ is the total number of particles of the dark halo
component, $\alpha$ a is model-dependent factor whose value must be set
empirically, and $c$ is the halo concentration inferred from the median
concentration-mass relation used in the galaxy halo models of \citet{DS10}
\begin{equation}\label{M-c}
c(M,z)=9.35\left[\frac{\Mvir}{10^{12}\;h^{-1}\msun}\right]^{-0.094}(1+z)^{-1}\;,
\end{equation}
which provides a resonable approximation to the results from high-resolution $N$-body simulations of the standard concordant $\Lambda$CDM cosmology measured over the ranges of halo masses, $M\sim 10^{12}M_\odot$, and redshifts, $z\sim 0$, we are interested in.\footnote{Recent investigatigations of the $c(M,z)$ relation \citep[see, e.g.\ the review by][and references therein]{Oko17} have confirmed that the evolution of the halo concentration is more complex than such a simple
form based on the inside-out growth of structure \citep[][]{Bul01}. However, most concentration-mass relations from the literature show a very good agreement with eq.~(\ref{M-c}) in the regime relevant for this work.}.
For our model galaxies, we find that to get the desired
amount of angular momentum we must take $\alpha\simeq 1/2$, in good
agreement with the radial velocity dispersion profile of isotropic NFW
haloes with $c\sim 10$ \citep{LM01}. We refer the reader to
Appendix~\ref{halorot} for the definitions of parameters $\lambda$,
$c$ and $\alpha$.

%Table 1
\begin{table*}
%\begin{minipage}{113mm}
%\scriptsize
\centering
%\rotate
\caption{Initial parameters$^a$ for major mergers with $r_{\rm cir,p}=4/3$}
\label{orb_params}
%\begin{threeparttable}
%\tablecolumns{8}
%\tablewidth{0pt}
%\tablenum{1}

\begin{tabular}{cccc|cccc}
\hline\hline

%\tabletypesize{\scriptsize}
$\eta$ & $\mu$ & $r_0$ & $v_0^2$ & $\epsilon$ & $r_{\rm per}$ &
$(v_{\rm x},v_{\rm y})_0$ & $\mid v_{\rm x,0}\mid : \mid v_{\rm
  y,0}\mid$ \\ \hline & & & & 0.20 & 0.0269 & $(-0.6880,0.1633)$ &
$\sim 4:1$ \\ 1 & 1/2 & 2 & 1/2 & 0.45 & 0.1426 & $(-0.6042,0.3674)$ &
$\sim 2:1$ \\ & & & & $0.70$ & 0.3905 & $(-0.4082,0.5774)$ & $\sim
1:1$ \\ \hline & & & & 0.20 & 0.0269 & $(-0.7329,0.1591)$ & $\sim 5:1$
\\ 2 & 1/3 & 16/9 & 9/16 & 0.45 & 0.1426 & $(-0.6591,0.3580)$ & $\sim
2:1$ \\ & & & & $0.70$ & 0.3905 & $(-0.4961,0.5625)$ & $\sim 1:1$
\\ \hline & & & & 0.20 & 0.0269 & $(-0.7579,0.1600)$ & $\sim 5:1$ \\ 3
& 1/4 & 5/3 & 3/5 & 0.45 & 0.1426 & $(-0.6859,0.3600)$ & $\sim 2:1$
\\ & & & & $0.70$ & 0.3905 & $(-0.5292,0.5657)$ & $\sim 1:1$ \\ \hline
\end{tabular}
\tablefoot{$^a$We provide values for: the ratio between the masses of the primary and secondary progenitors, $\eta$; the reduced mass of the system, $\mu$; the initial modulus of the relative position vector of the galaxy centres and its time derivative, $r_0$ and $v_0$; the initial orbital circularity, $\epsilon$; and the initial pericentric distance $r_{\rm per}$. All mergers take place on the $XY$ plane, with
  the orbital spin lying along the $Z+$ axis and the relative position
  vector initially directed along the $X+$ axis, so $v_{\rm x,0}$ and $v_{\rm y,0}$ are, respectively, the radial and tangential components of the initial
  relative velocity.  The principal axes of
  the galaxies are initially aligned with the Cartesian axes. All
  dimensional quantities are given in simulation units (see
  Section~\ref{models}).}
%\end{threeparttable}
%\end{minipage}
\end{table*}

%& & & & 0.20 & 0.0269 & $(-0.7739,0.1614)$ & $\sim 5:1$ \\
%4 & 1/5 & 8/5 & 5/8 & 0.45 & 0.1426 & $(-0.7023,0.3631)$ & $\sim 2:1$ \\
%& & & & $0.70$ & 0.3905 & $(-0.5472,0.5705)$ & $\sim 1:1$ \\
%\hline

\subsection{Initialization of the models}\label{initialization}

Traditionally, idealised binary merger simulations had to face the
lack of knowledge of the appropriate initial conditions. In recent
years, however, several high-resolution, large-scale cosmological
simulations have carried detailed investigations of the orbital
parameters of merging pairs of CDM haloes providing sufficient
information to fill this gap \citep[e.g.][]{Ben05,KB06,McCa12}. In the
present work, the full set-up of the mergers is based on cosmologically
motivated initial conditions. This means that the values of all the
parameters that define the initial orbital configurations of the pairs
are set taking into account their corresponding probability
distribution functions (PDF) predicted by currently favoured
cosmological models.

\subsubsection{Parameters that define the orbits}\label{ini_cond}

We have set the initial orbital parameters of our merging galaxies by assuming that at $t=0$ they behave to a large extent as point-mass objects, which allows for the
application of the solution of the Keplerian two-body problem.

As discussed in Appendix~\ref{one-body eqs}, under this approximation three parameters are required to completely define the initial orbit of two interacting galaxies. For the present investigation, we
have chosen the initial modulus $r_0$ of the relative position vector
connecting the nuclear regions of the galaxies, which in all our
experiments is taken equal to the sum of the two virial radii of the
galactic halos\footnote{This choice leads to initial values for its time derivative, $v_0$, i.e.\ the relative radial velocity of the galaxies, $\sim 100$--$150$ \kms\ that are consistent with the observed 'cold Hubble flow' (low value for the root-mean-square galaxy velocity) in the Local Volume \citep*[see, e.g.][and references therein]{TCB05}.}, the initial orbital energy, which we represent, as usual, by the dimensionless parameter $r_{\rm cir,p}$,
expressing the radius of a circular orbit with the same orbital energy
of the merger in units of the virial radius of the primary halo, and
the initial orbital circularity, $\epsilon$, which is a proxy for the
initial orbital angular momentum.

Various authors \citep{Tor97,Zen05,KB06,Jia08,McCa12} have used
cosmological simulations of merging DM haloes to study the
distribution of orbital circularities in merging pairs. There is
general consensus that the circularity parameter follows a universal
distribution, with a central peak near $\epsilon\simeq 0.45$
(corresponding to an ellipticity $e\simeq 0.9$) and a strongly
platykurtic behaviour characterized by a substantial, log-normal
distributed spread with $\sigma\sim 0.4$, which accounts for the
stochastic nature of merger processes with equivalent initial
conditions, and that points to a relative overabundance of highly
circular bound orbits with respect to highly radial ones.  Based on
these results, and to take into account the impact that different
amounts of orbital angular momentum transfer may have on the remnants,
we will simulate collisions along three different orbital trajectories
defined by $\epsilon\simeq 0.20$,~$0.45$, and $0.70$. From now on, we
will refer to the geometry of these orbits, respectively, as radial,
tangential and elliptical.

As shown by \citet{McCa12}, the initial orbital energy of their
bias-free dataset obeys a right-skewed distribution which peaks
slightly beyond one and has a long tail towards extremely energetic
orbits. Following the results from these authors (see their Fig.\ 3),
and in line with what we are doing with the other orbital parameters,
we adopt three different values for $r_{\rm cir,p}$ that encompass the
bulk of \citeauthor{McCa12}'s distribution: $1.3\simeq 4/3$, intended
to approximate the modal value of their PDF, $2$, and $2.7\simeq 8/3$.

Table~\ref{orb_params} summarizes the values of the initial orbital
parameters we have adopted for galaxy pairs with a reduced initial
orbital energy $r_{\rm cir,p}=4/3$ and the three integer mass ratios
that commonly define major mergers. The Greek letter $\eta\equiv
M_{\rm p}/M_{\rm s}$ will be used henceforth to express the ratio
between the masses of the largest galaxy and its merger partner.

\subsubsection{Magnitude and relative orientation of internal spins}\label{spins}

Cosmological $N$-body simulations and observational estimates agree in
that the dimensionless spin parameter of galactic haloes (see equation
\ref{spin_param}) follows a roughly universal positively skewed
lognormal PDF, $P(\lambda)$, of median, $\lambda_0$, close to $0.04$,
even when the non-gravitational processes acting on the baryons are
considered \citetext{see, for instance, \citealt{Sha06,Her07}, and
  \citealt{Bry13} and references therein}. In an attempt to take into
account the wide range of values shown by this parameter, we have
decided to construct late-type galaxy models which include up to four
different dark halo spin values: $\lambda=0.02$, $0.04$, and $0.06$,
which sample the central region of $P(\lambda)$ where the probability
is highest, as well as $\lambda=0.0$, which is used as a reference for
calibrating the impact of halo rotation in our results. In the case of
early-type galaxies, equation~(\ref{fh}) with $\alpha=1/2$ is simply
regarded as a convenient formula which allows one to provide haloes
with the right spin. Since it is reasonable to expect that the dark
haloes associated with early-type galaxies have less angular momentum
than those associated with discs, we have modelled the former using
solely $\lambda=0.02$.

The spin of galactic haloes can have a substantial impact on the
encounter rate of their constituent particles, affecting the strength
of the dynamical friction force mutually exerted by the galaxies and,
by extension, the merger time (see Section~\ref{parameters}) and even
the structure of the remnant. Therefore, the outcome of a galaxy
collision is expected to depend on both the moduli and the degree of
alignment between the orbital and haloes spin vectors. To assess the
full scope of this contribution, besides simulating haloes with
different rotation speeds (see Appendix~\ref{halorot}), we have
further adopted twelve extremal configurations for the initial
relative orientations of the galaxies that maximize/minimize the
coupling between the internal and orbital spin
vectors\footnote{According to the numerical investigations of merging
  by \citet{Ben05} and \citet{KB06} the angles between the two spin
  planes of the galaxies, between each one of the spin planes and the
  orbital plane, between the infall direction and the infall velocity,
  and between the infall velocity of satellites and the angular
  momentum of their host haloes are all essentially
  uncorrelated.}. Given that the orbits of all our mergers lie in the
$XY$ plane, we can use the standard versors of the 3-D Cartesian
coordinate system, $\hat{\textbf{\i}}$, $\hat{\textbf{\j}}$, and
$\hat{\bf k}$, to identify the different directions initially adopted
for the internal spins of the primary and secondary galaxies, ${\bf
  J}_{\rm p}$ and ${\bf J}_{\rm s}$, respectively. They are listed in
Table~\ref{halo_spins}.

\subsection{The library of binary mergers}\label{library}

The evolution of the galaxy mergers has been performed using
GyrfalcON, a full-fledged serial $N$-body tree-code using
\citeauthor{Deh00}'s force algorithm of complexity $\cal{O}(N)$
\citep{Deh00} implemented in the \nemo\ toolbox. We take $N_{\rm
  crit}=6$ for the maximum number of bodies that avoid cell splitting
when building the tree, a tolerance $\theta=0.40$, as well as a
longest timestep of $\sim 0.001$ simulation time units. The time
integration adopted uses an adaptive block-step scheme which assigns
individual time-steps to the bodies by dividing the longest step onto
$2^6$ parts as controlled by the parameters {\texttt fac}$\ =0.03$ and
{\texttt fph}$\ =0.0015$. For all simulations the merger remnants were
allowed to evolve $\sim 1.5\;\tsu$ ($\sim 3$ Gyr) after the
coalescence of the centres of the two galaxies. This has enabled us to
obtain as a byproduct a massive suite of merger remnants suitable, for
instance, to study the impact of initial conditions on the structure
of these systems and the manner they approach their final state of
equilibrium. This and other related questions will be addressed in
forthcoming papers.

We have tested numerical convergence in our merging pair simulations
by running a specific subset of configurations: DD equal-mass mergers
with modal values of the reduced orbital energy and halo spin
parameter ($4/3$ and $0.04$, respectively), but with a factor five
higher resolution ($N$). Comparison of the intercentre separation
evolutions shows negligible differences, except for the final
pericentric passage and merger times, which on average are shorter at
high resolution by only a small $1.3\%$ or about 60 Myr. We take this
value as indicative of the typical uncertainty in the inferred merger
times. The primary data output is made every $0.015$ simulation time
units or $\sim 30$ Myr.

%Table 2
\begin{table}
%\scriptsize
\centering
%\rotate
\caption{Initial orientations of the internal spins of the primary (p)
  and secondary (s) progenitors}
\label{halo_spins}
%\begin{threeparttable}
%\tablecolumns{3}
%\tablewidth{0pt}
%\tablenum{2}

\begin{tabular}{lcc}
\hline\hline

ID code$^a$ & direction of ${\bf J}_{\rm p}$ & direction of ${\bf J}_{\rm s}$ \\
\hline

DD & $+\hat{\bf k}$ & $+\hat{\bf k}$ \\
DR & $+\hat{\bf k}$ & $-\hat{\bf k}$ \\
RR & $-\hat{\bf k}$ & $-\hat{\bf k}$ \\
DX90 & $+\hat{\bf k}$ & $-\hat{\textbf{\j}}$ \\
DY90 & $+\hat{\bf k}$ & $+\hat{\textbf{\i}}$ \\
RX90 & $-\hat{\bf k}$ & $-\hat{\textbf{\j}}$ \\
RY90 & $-\hat{\bf k}$ & $+\hat{\textbf{\i}}$ \\
X90X90 & $-\hat{\textbf{\j}}$ & $-\hat{\textbf{\j}}$ \\
X90X270 & $-\hat{\textbf{\j}}$ & $+\hat{\textbf{\j}}$ \\
Y90Y90 & $+\hat{\textbf{\i}}$ & $+\hat{\textbf{\i}}$ \\
Y90Y270 & $+\hat{\textbf{\i}}$ & $-\hat{\textbf{\i}}$ \\
X90Y90 & $-\hat{\textbf{\j}}$ & $+\hat{\textbf{\i}}$ \\
\hline
\end{tabular}
\tablefoot{$^a$D stands for 'Direct', i.e.\ parallel to the orbital
  spin which in all cases is oriented along the $Z+$ axis, R for
  'Retrograde', i.e.\ anti-parallel to the orbital spin, X90 means the
  galaxy is rotated 90 degrees counterclockwise around the $X$ axis,
  Y90 represents a positive rotation of 90 degrees around the $Y$
  axis, etcetera.}
\end{table}

As described in the previous sections, the suite of binary mergers investigated in this paper combines different realizations of six quantities: mass ratio, orbital energy, orbital eccentricity (or circularity), module and orientation of the halo spin, and galaxy morphology. Nevertheless, not all the combinations allowed by the subsets of discrete values adopted for them have been performed. In an effort to optimize memory space and computational time, we have avoided certain combinations of parameters that add
nothing new to the investigation. Thus, the core of the encounter survey, in
which most of the results of this study are based on, is formed by 144
Sb+Sb ($\eta=1$) mergers -- this figure arises from multiplying the
three values adopted for the circularity of orbits by the four
possible values of the spin of galactic halos (in our simulations the
two progenitors share identical values of $\lambda$) and by the twelve
distinct relative orientations of these spins --, 144 Sb+Sc ($\eta=3$)
mergers, and 36 E3+E3 ($\eta=1$) mergers, all run with a reduced
orbital energy $r_{\rm cir,p}$ of $4/3$. This primary sample is
completed with 144 Sb+Sb ($\eta=1$) mergers, plus 36 E3+E3 ($\eta=1$)
mergers with orbital energy $2.0$, as well as by 36 E3+Sc ($\eta=3$)
and 36 Sb+E3 ($\eta=3$) mergers with $r_{\rm cir,p}=4/3$. All subsets
with an E3 galaxy have $\lambda$ fixed to 0.02). Finally, a subset of
18 additional S+S encounters with $r_{\rm cir,p}$ of $8/3$ and
$\lambda=0.04$, are run with the sole purpose of completing a number
of combinations of orbital parameters which allow a fitting formula
for major mergers to be obtained with a minimum level of accuracy (see
Section~\ref{tau_mer_fits}). This gives a total of 594 major merger
simulations, which is the largest science sample of numerical models
of isolated galaxy pairs ever built to study this type of phenomena.

\section{Merging time-scale and tracking scheme}\label{tau_merge}

Our definition of the merger time-scale, $\tau_{\rm mer}$, follows the
same general lines adopted in previous numerical studies of dynamical
friction dealing with galaxy mergers either isolated or developed
within a cosmological context
\citep[e.g.][]{BKMQ08,Jia08,McCa12,Vil13}. The start of a merger is
defined as the instant when the centre of mass of the satellite-to-be
galaxy first crosses the virial radius of its host halo \citep{Ben10}
-- so the initial radial distance between a given primary galaxy and
its satellites is always the same --, while a merger ends at the
instant when it is no longer possible to distinguish that we are
dealing with two separated entities. There are, however, several ways
to pinpoint the endpoint which yield similar but not identical
results. A popular criterion based on the classically perceived
picture of the merger process is to use the time at which the
satellite has lost (almost) all of its specific angular momentum
relative to the host \citep[e.g.][]{BKMQ08}. Alternatively, one can
seek for the minimization of some other function of the relative
separation of galaxies in phase space $\Delta (t)\equiv f(r(t),v(t))$
or of the number of bound particles of the satellite, $N_{\rm
  bound,s}(t)$. This last option is best suited to track mergers in
which satellites are fully tidally disrupted before the galaxies
coalesce, which is not the case in our major merger
simulations. Approaches of this kind, based on the nullification of
some characteristic time-decreasing property, have to face a serious
difficulty: it is computationally unfeasible to achieve such
cancellation. To overcome this problem, a frequent recourse -- used,
for example, in the works cited at the beginning of this section -- is
to adopt a priori a reasonable minimum threshold for said property,
generally expressed as a small fraction of its initial value (a $5\%$
is the most common choice), below which the merger is considered as
complete. Nevertheless, there is no guarantee that this procedure
works well in all situations. To begin with, the fact that none of the
properties suitable for tracking the evolution of a merger follows an
entirely monotonic decrement entails the risk that any threshold,
however small it is, is reached occasionally before the end of the
merger, which would then lead to underestimate the merger
time-scale. Besides, the adoption of an extremely low threshold value
is not the solution either, since the aim is defining a standard
metric that can be applied to any simulation regardless of its
resolution. This has led us to propose a new strategy based on the
monitoring of the temporary evolution of the behaviour, rather than
the quantitative value, of $\Delta (t)$. As we show below, the shape
of the intercentre separation history provides not only a
straightforward but, more importantly, an accurate and resolution-free
way of defining the instant at which the galaxies coalesce, especially
for the type of simulations we are dealing with.

%Fig. 1
\begin{figure}
	\centering
	\resizebox{\hsize}{!}{\includegraphics[keepaspectratio,clip=true,angle=0]{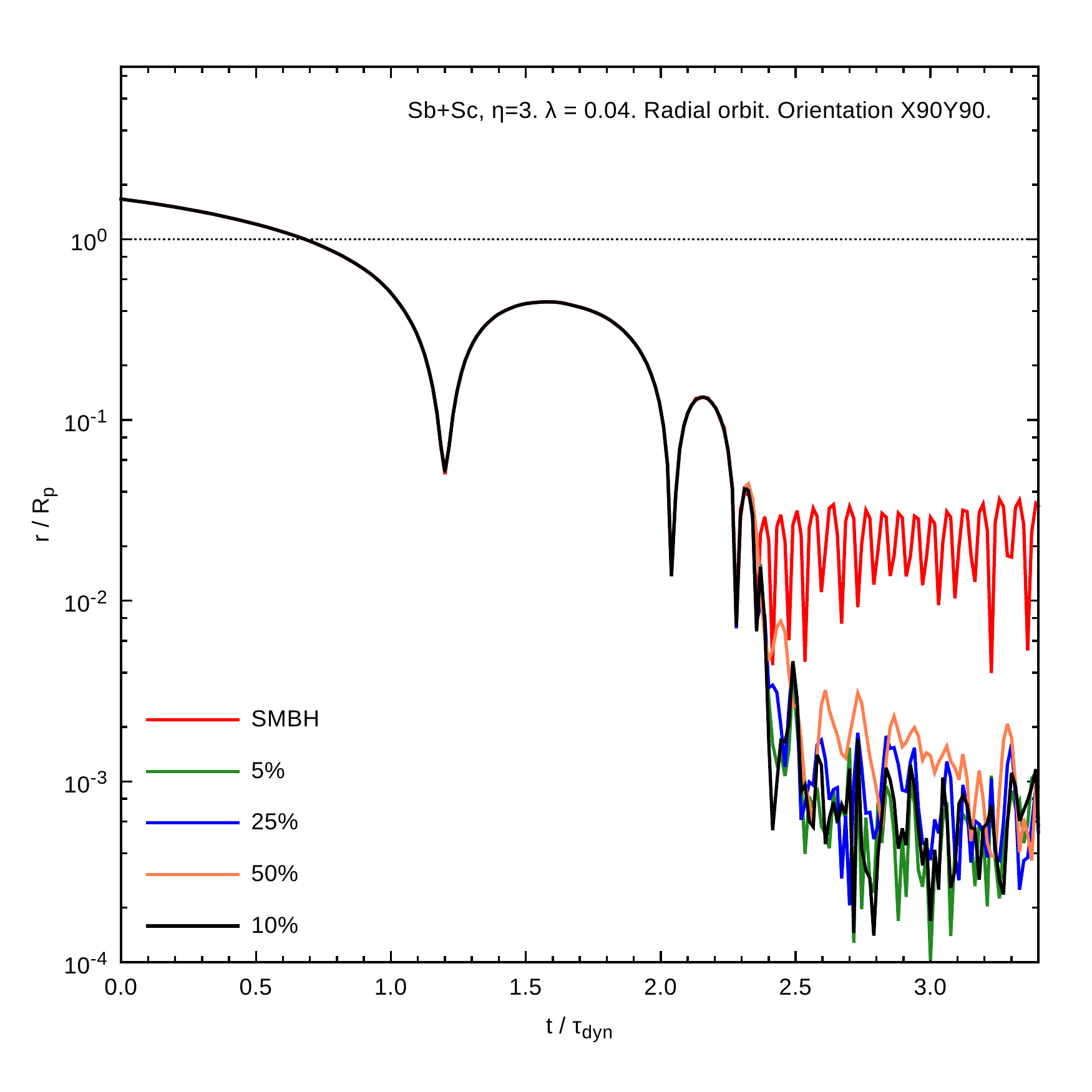}}
	\vspace*{-10mm}
	\caption{\small Time evolution of the intercentre distance $r$, in
		units of the radius $R_{\rm p}$ of the primary halo, of a pair of
		Sb+Sc galaxies with mass ratio 3:1 and internal spin parameter
		$\lambda=0.04$ colliding along a radial orbit. The initial
		orientation of the internal spin of the haloes is X90Y90 (see
		Table~\ref{halo_spins}). Different solid curves show merger
		histories calculated using different definitions of the galactic
		cores: green is for the $5\%$ fraction of the most bound luminous
		particles, black for the $10\%$, blue for the $25\%$, coral for the
		$50\%$, while the red line shows the merger evolution as determined
		from the central SMBH. The intersection of these
		curves with the gray horizontal dotted line marks the onset of the
		merger. The plot shows that the final coalescence of the
		galaxies at $t/\tau_{\rm dyn}\sim 2.4$ is better delineated when the
		galactic cores are defined using a moderate fraction, between 5 and
		25 per cent, of the most bound particles. }\label{fig_1}
\end{figure}

%Fig. 2
\begin{figure*}
	\centering
	\includegraphics[width=\linewidth,keepaspectratio,clip=true,angle=0]{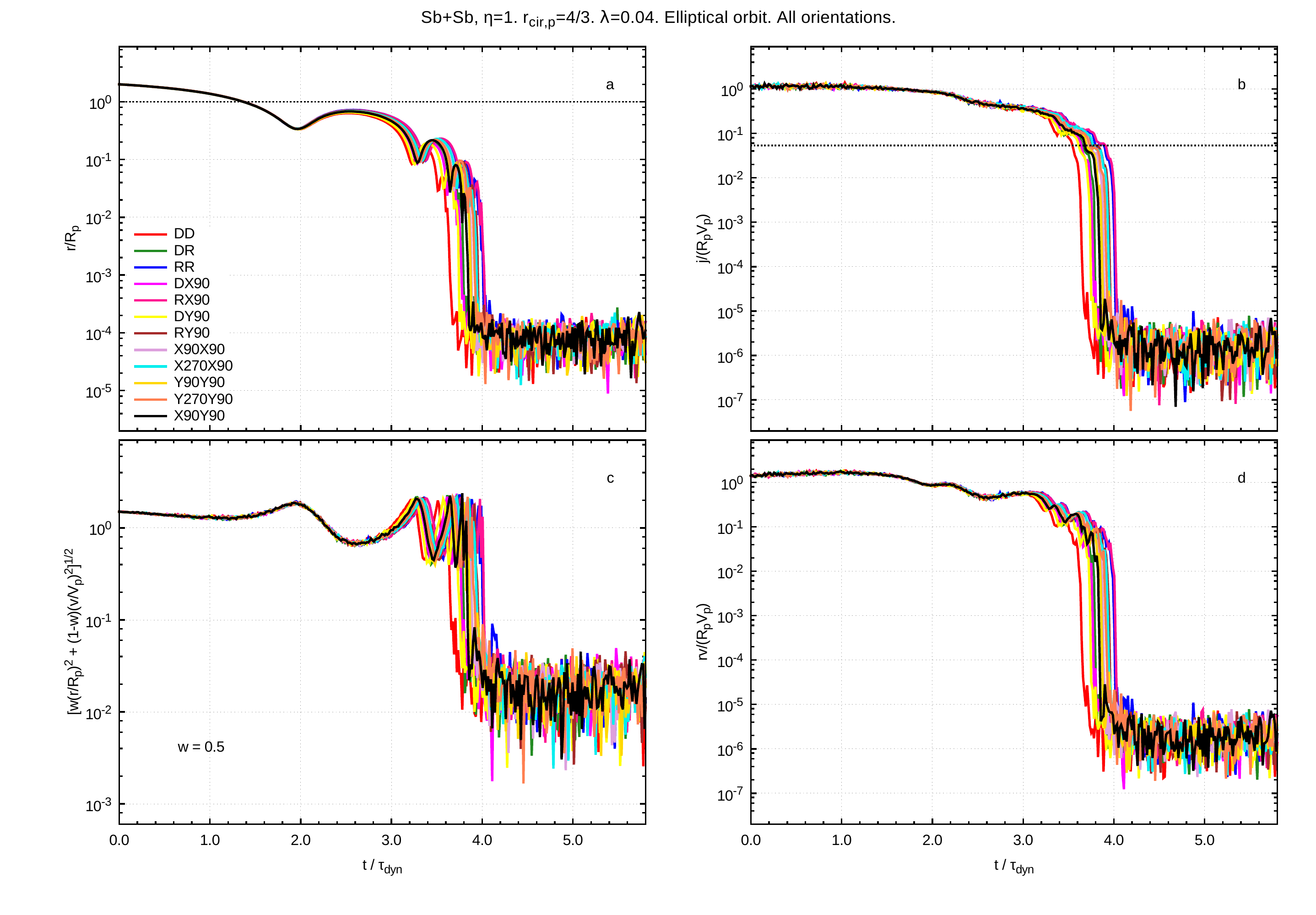}
	\vspace*{-5mm}
	\caption{\small Time evolution of the intercentre separation of Sb+Sb
		mergers of equal-mass galaxies with internal spin parameter
		$\lambda=0.04$ colliding along elliptical orbits. Different panels
		show different choices for $\Delta (t)$: (a) the intercentre
		distance; (b) the specific orbital angular momentum; (c) a weighted
		non-canonical distance in phase space based on the Euclidean notion
		of distance; (d) the product of the intercentre distance and
		relative velocity. Normalization units $R_{\rm p}$ and $V_{\rm
			p}$ refer, respectively, to the virial radius and velocity of the
		primary halo. Galaxy centres are calculated from the $10\%$ fraction
		of the most bound luminous particles. Coloured solid curves in
		each panel show the merger histories corresponding to the twelve
		initial orientations adopted for the internal spin of the haloes
		(see legend in panel a and Table~\ref{halo_spins}). In panel (a),
		the intersection of the curves with the black horizontal dotted line
		marks the onset of the mergers. In panel (b), the black horizontal
		dotted line marks the value of $j_{05}$ inferred for the black solid
		curve (X90Y90). The values of $j_{05}$ for the other curves are very
		similar.}\label{fig_2}
\end{figure*}

%Fig. 3
\begin{figure*}
	\centering
	\includegraphics[width=\linewidth,keepaspectratio,clip=true,angle=0]{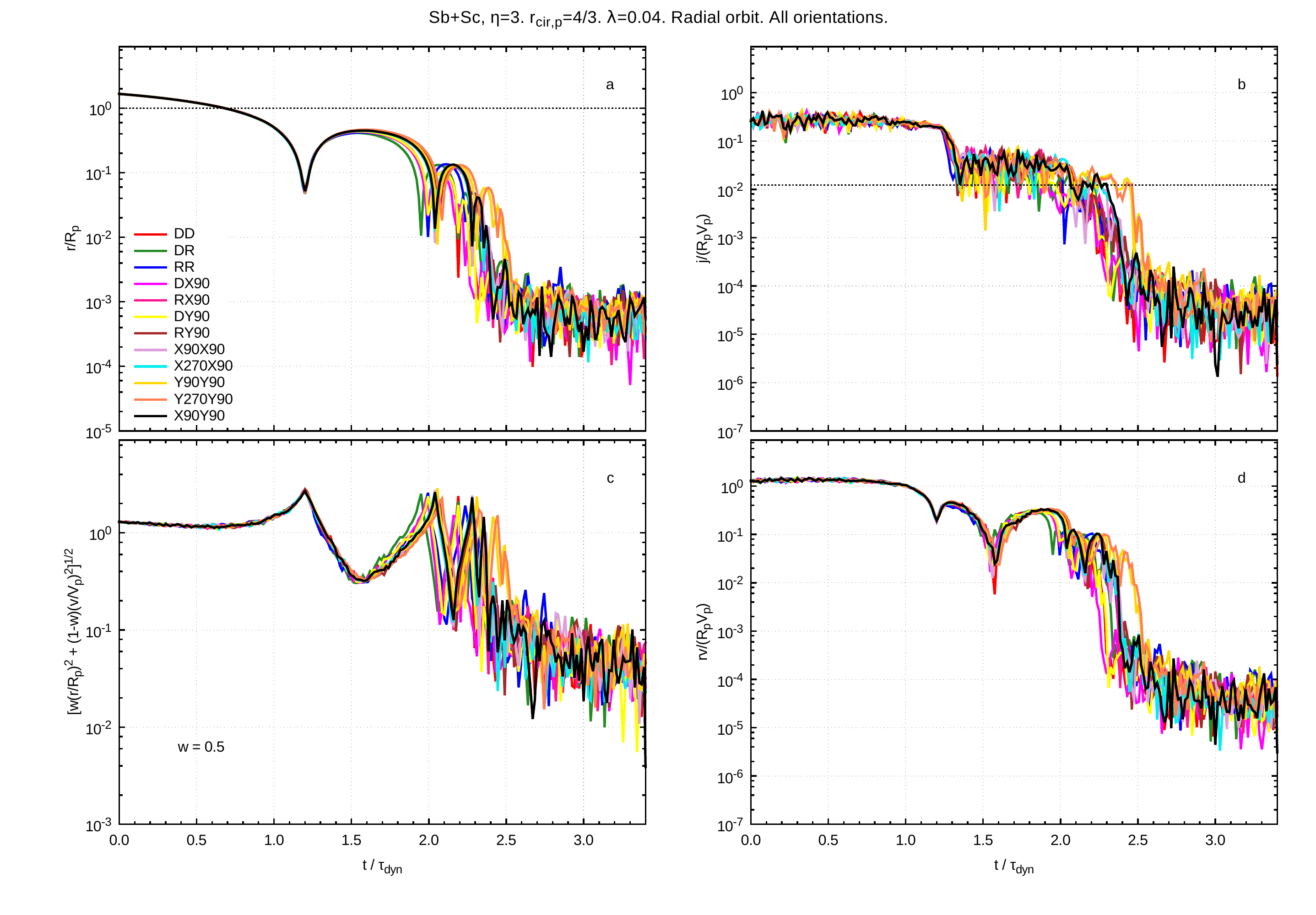}
	\vspace*{-5mm}
	\caption{\small Same as Fig.~\ref{fig_2} but for Sb+Sc mergers with
		mass ratio 3:1 colliding along radial orbits.}\label{fig_3}
\end{figure*}

To track the orbital evolution of the galaxies throughout the span of
the entire simulation we have reduced, in each snapshot, the galactic
haloes to a system of two point masses. Thus, throughout the entire
simulations, the progenitors are represented in phase-space by the six
coordinates of the centre of mass of the upper decile of the most
bound particles of their baryonic cores, which constitute the region
of highest matter density for any galaxy\footnote{Since the galactic
  cores have a larger spread in velocity than in position, a 10th
  percentile of the stars of each galaxy averages over enough bodies
  to provide an accurate determination of the galaxy velocity. This is
  also the reason why all variables that depend on $v(t)$, such as the
  orbital angular momentum, present a larger level of noisiness than
  those depending on $r(t)$.}. Particle binding energies are computed
according to the gravitational potential calculated from all the
remaining bodies assigned to the parent galaxy at the start of the
merger. We show in Fig.~\ref{fig_1} the time evolution of the mean
separation in configuration space between the dynamical centres,
defined using different fractions of bound particles as well as by the
particle representing the central SMBH, of a pair of Sb+Sc galaxies
colliding along a radial orbit with an initial energy $r_{\rm cir,p}$
of $4/3$, so in this case $\Delta (t)=r(t)$. It is obvious from this
plot that galactic cores defined using the average position of subsets
of the most bound particles allow one to track the evolution of the
merger even along its most advanced stages, when the global structure
of the progenitors is heavily disrupted by the violently changing
gravitational field and the variation of their overall centre-of-mass
positions and velocities is not predictable anymore. As this evolution
is computed, it is straightforward to identify the instants $t_i$ of
closest approach (pericentric passages) as those points in which the
relative distance between the dynamical centres of host and satellite,
$r(t)$, reaches a local minimum (that behaves as a non-differentiable
type I discontinuity). Besides, the logarithmic scale used in the $Y$
axis permits appreciation of the fact that coalescence is punctuated
by a nearly instantaneous and, in most cases,
sharper-than-a-pericentric-passage drop in the intercentre distance
that resembles a jump or step discontinuity (in fact we shall see that
this abrupt minimization is extensive, to a greater or lesser extent,
to any function of the six coordinates of phase space). Due to the
presence of this well-defined step, it is easier in practice to
identify the time of coalescence, $t_{\rm coal}$, by starting the
analysis of the time history of $\Delta(t)$ beyond the completion of
each merger and tracing back its evolution. The coalescence of the
galaxies takes place when the relative variation of the separation
reaches its largest value, which is reflected, for example, in the
maximization of the time derivative of the logarithm of $\Delta
(t)$. To facilitate the identification of such jump, sometimes it may
be necessary to apply a low-pass filter to $\Delta (t)$.

In addition, Fig.~\ref{fig_1} reveals that shortly after coalescence
the intercentre separation, still calculated from the bodies initially
assigned to each parent galaxy, begins to oscillate steadily around a
constant residual value with an also constant amplitude -- the fact
that the particles defining the galactic cores can vary between
consecutive steps explains why these fluctuations are not fully
harmonic. We note that, when the dynamic centres are optimally
defined, the mean intercentre distance in the post-coalescence phase
provides a direct estimate of the minimum spatial resolution scale of
the remnant. In our 3:1 mergers, it is found that this value is $\sim
5\times 10^{-4}\rsu$ (see Fig.~\ref{fig_1}), while for 1:1 mergers
decreases until reaching $\sim 10^{-4}\rsu$, which just happens to be
comparable to the force softening scale adopted for the luminous
component. This implies that our choice of softening length provides a
good compromise between efficiency and fidelity for the present
$N$-body experiments.

\subsection{An accurate metric for major mergers}\label{optimal_marker}

In Figs.~\ref{fig_2} and \ref{fig_3} we show four examples of
functions that can be used to track the merger history $\Delta (t)$ in
two extreme encounter configurations. Fig.~\ref{fig_2} depicts the
merger of equal-mass disc galaxies along elliptical orbits, while
Fig. \ref{fig_3} shows the evolution of Sb+Sc 3:1 mergers along radial
orbits, in both cases with $\lambda =0.04$. The twelve curves shown in each
panel correspond to the twelve extremal relative orientations of the
galaxies defined in Table \ref{halo_spins}. A comparison between both
Figures shows some common trends as well as noticeable
differences. Among the former it is the fact that for the two groups
of merger histories the different relative orientation of the discs
leads to appreciable variations among the times of coalescence, in the
order of $0.3$--$0.4\;\tsu$ or $\sim 0.6$--$0.8$ Gyr. On the other
hand, the main differences have to do with the fact that collisions
along the most circular orbital configurations proceed in a more
orderly fashion, wherein it is hinted that the strength of the
dynamical friction, i.e.\ the duration of the mergers is modulated by
the initial relative orientation of the internal spin of the principal
halo with respect to the orbital spin (see also next
section). However, this is not the case for radial collisions, where
we observe that the maximum length of mergers corresponds to the
'Y90Y90' and 'Y270Y90' configurations. This indicates that the more
eccentric the orbits the more effectively collisions scramble the
initial internal spin of the merging haloes.

Pericentric -- and, depending on the function used to estimate
$\Delta(t)$, apocentric -- passages are also more clearly punctuated
in head-on collisions. The most marked differences correspond to the
merger histories depicted on panels b and d. The specific orbital
angular momentum $j(t)$ represented in panels b starts experiencing
noticeable losses only after the first pericentric passage, but its
evolution is essentially monotonic and nearly constant in time for
elliptical orbits, while it is impulsive (step-wise) for radial
collisions, at least along the firsts pericentric passages. In
contrast, the evolution of the product of the relative distance and
velocity of the centres, $rv(t)$, shown in panels d is always
non-monotonic, with the most eccentric collisions showing
substantially deeper local minima associated with passages through the
extreme points of the orbit.

%Fig. 4
\begin{figure*}
	\centering
	\resizebox{\hsize}{!}
	{\includegraphics[width=\linewidth,keepaspectratio,clip=true,angle=0]{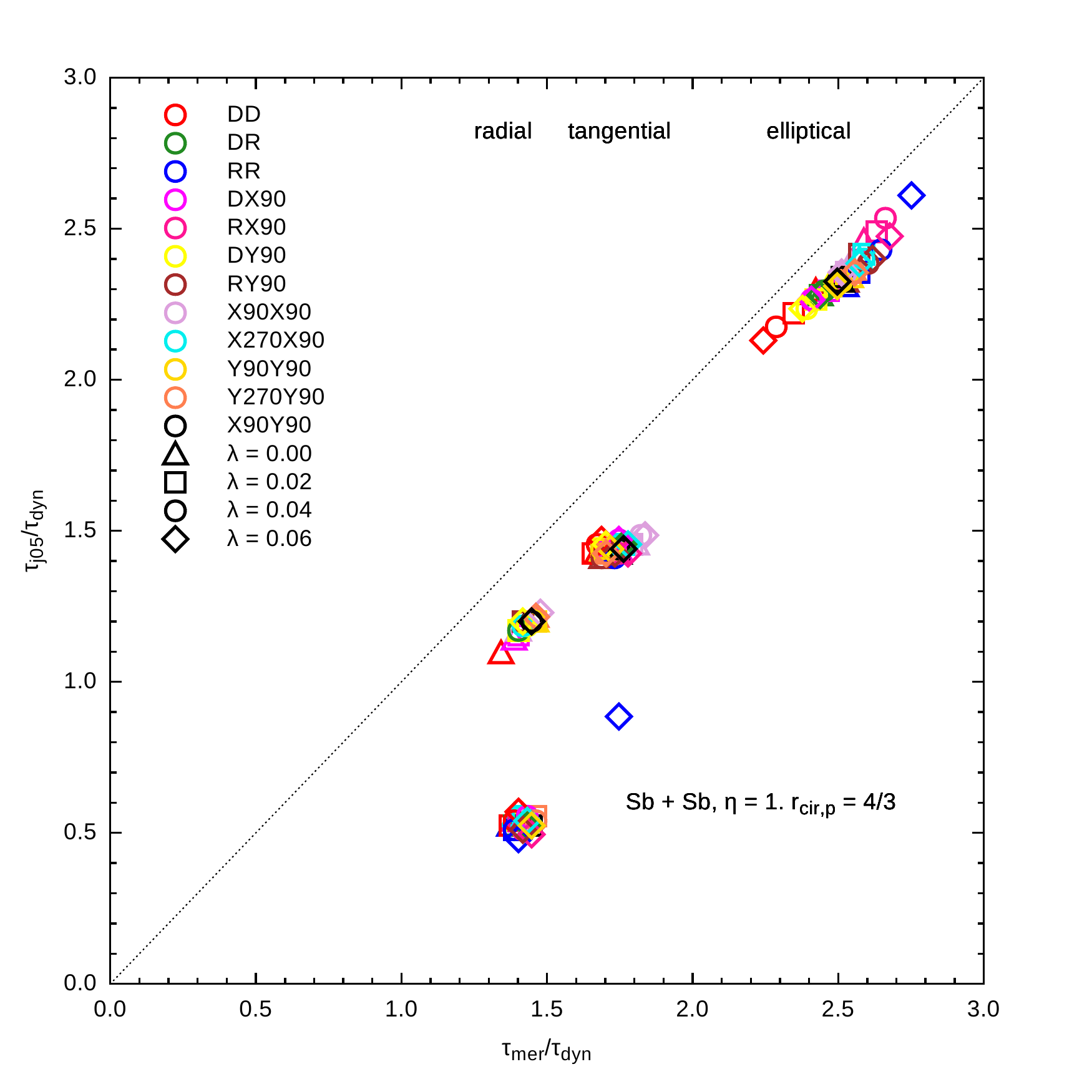} 
	\includegraphics[width=\linewidth,keepaspectratio,clip=true,angle=0]{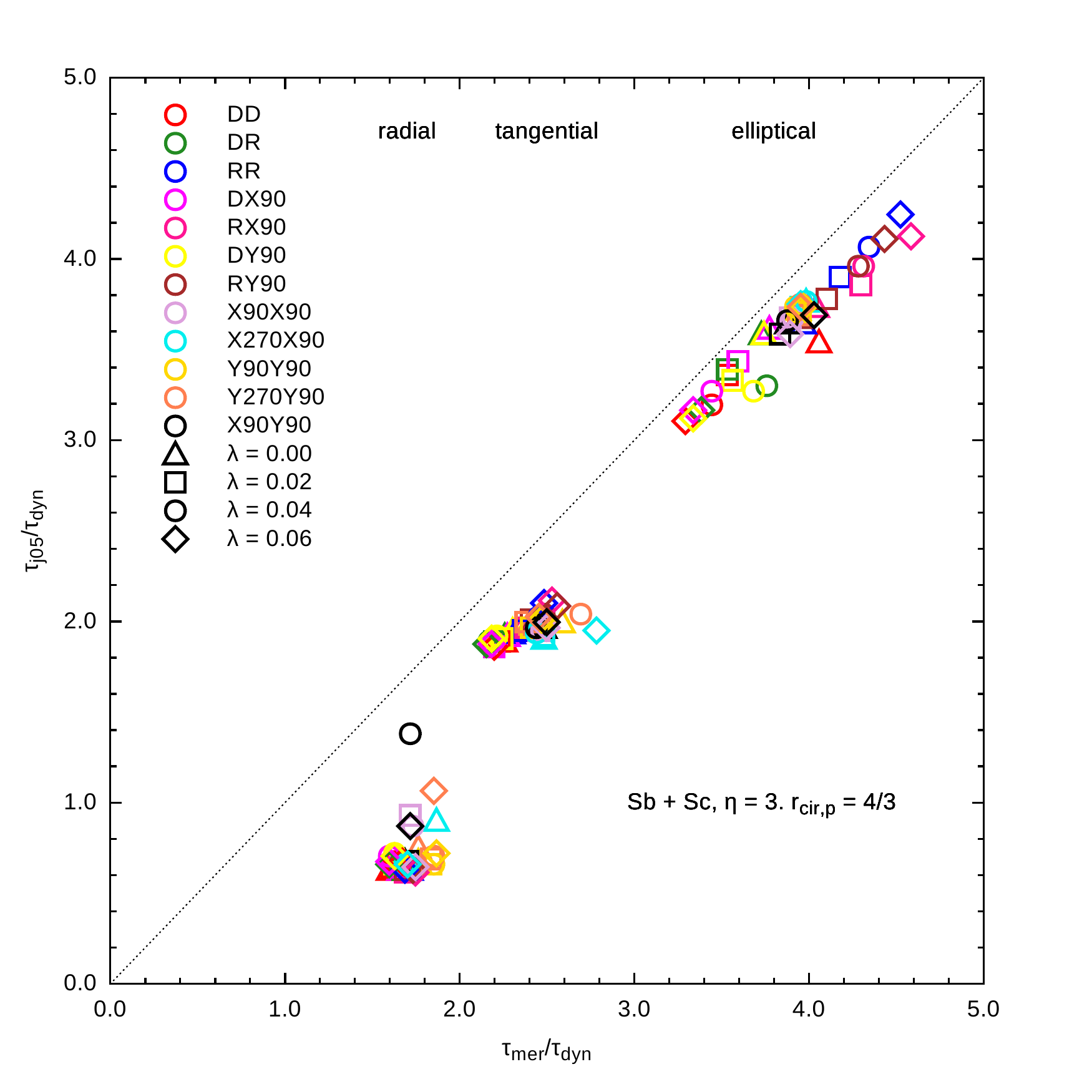}}
	\caption{\small Comparison between $\tau_{\rm mer}$ and
		$\tau_{j05}$ measured from all the mergers depicted in
		Figs.~\ref{fig_2} (left) and \ref{fig_3} (right). In
		all cases $\tau_{l05}<\tau_{\rm mer}$, even when $j(t)$ behaves
		as a pure monotonically decreasing function.}
	\label{fig_4}
\end{figure*}

Fig.~\ref{fig_2} and, especially, Fig.~\ref{fig_3} also reveal that,
whatever merger tracker function is chosen, the sinking of the centres
can continue for quite a long time after $\Delta (t)$ has dipped
within five per cent of the value it had the first time the satellite
galaxy crossed the virial radius of the primary (this threshold for
$j(t)$, which we refer to as $j_{05}$, is indicated by the horizontal
black dotted line in the b panels). It is therefore apparent that
definitions of the merger time-scale based on this or similar
conditions may lead to the systematic underestimation of the length of
the merger. Thus, for the particular case of $j_{05}$, our
Fig.~\ref{fig_4} indicates that, regardless of the mass ratio of the
progenitors, the use of this quantity to calculate merger time-scales
results in fractional differences with a median value of about $6\%$
for major mergers along elliptical orbits. The differences raise to
$\sim 16\%$ at intermediate orbital circularities and distribute into
two narrow and clearly separated peaks around $20\%$ and $60\%$ for
radial collisions (in the bottom panel of Fig.~\ref{fig_4},
corresponding to the Sb+Sc case, most data points exceed by far the
$20\%$). Besides, comparison of Figs.~\ref{fig_2} and \ref{fig_3},
shows that the difficulty in monitoring the final act of the orbital
decay increases with increasing mass ratio and orbital eccentricity,
which entails that not all possible merger histories $\Delta (t)$ are
equally suitable to determine $t_{\rm coal}$. Among those investigated
in Figs.~\ref{fig_2} and \ref{fig_3}, the best overall behaviour
corresponds to the product $r(t)v(t)$ although the difference
vis-\`a-vis with $j(t)$ is very small -- this makes sense because
${\bf l}$ is simply the cross product of ${\bf r}$ and ${\bf v}$, so
its magnitude carries only an extra $\sin\theta$ factor. Nevertheless,
since the numerical calculation of the former function is not only
straightforward but somewhat more stable numerically, we adopt the
secular evolution of $rv$ as our preferred merger tracker.

\section{Dependencies of ${\mathbf\tau}_{\bf mer}$}\label{parameters}

%Fig. 5
\begin{figure*}
	\centering
	\includegraphics[width=\linewidth,keepaspectratio,clip=true,angle=0]{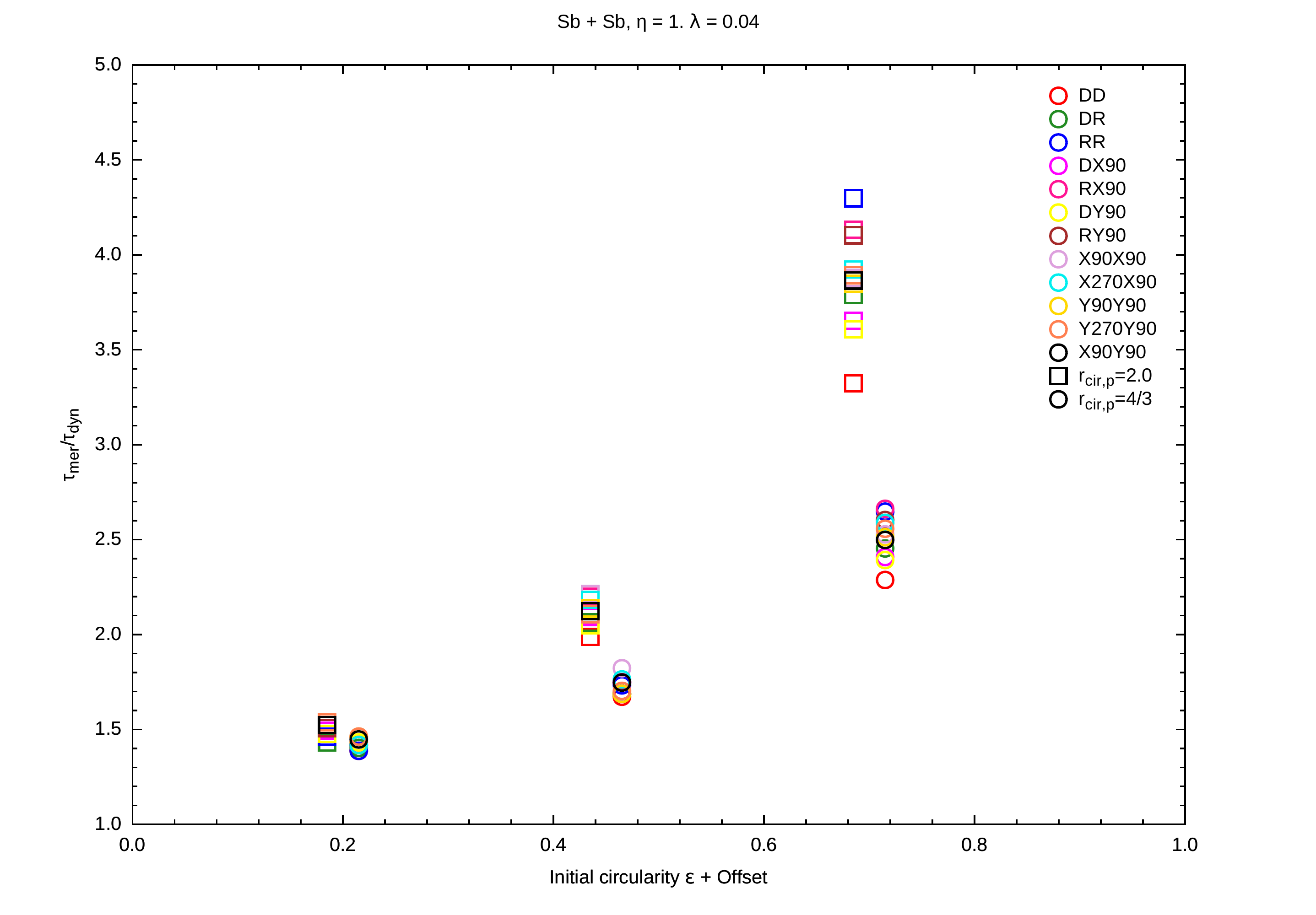}
	\caption{\small Examples of merger times measured in our simulations,
		$\tau_{\rm mer}$, as a function of the initial circularity of the
		orbit, $\epsilon$, showing the effect of varying the orbital energy,
		$r_{\rm cir,p}$, while keeping all other parametres unchanged. This
		plot shows results for equal-mass mergers of spiral galaxies with
		the modal value of the spin parameter ($\lambda=0.04$). Open squares
		are used for mergers with $r_{\rm cir,p}=2.0$, while open circles
		identify $r_{\rm cir,p}=4/3$ mergers. Data points are shown with
		different offsets around the true values of $\epsilon$ for
		clarity.}\label{fig_5}
\end{figure*}

In this section, we complete what has been said about the parameters
that control the length of mergers by validating not only the
dependencies expected from the classically perceived picture of this
phenomenon, i.e.\ variations with the initial circularity of the
orbits, orbital energy, and mass ratio, but also some others seldom
discussed in previous works -- at least with the level of detail that
we provide here --, such as dependencies on the magnitude and
orientation of the initial internal spins and on the morphology of the
progenitor galaxies.

%Fig. 6
\begin{figure*}
	\centering
	\includegraphics[width=\linewidth,keepaspectratio,clip=true,angle=0]{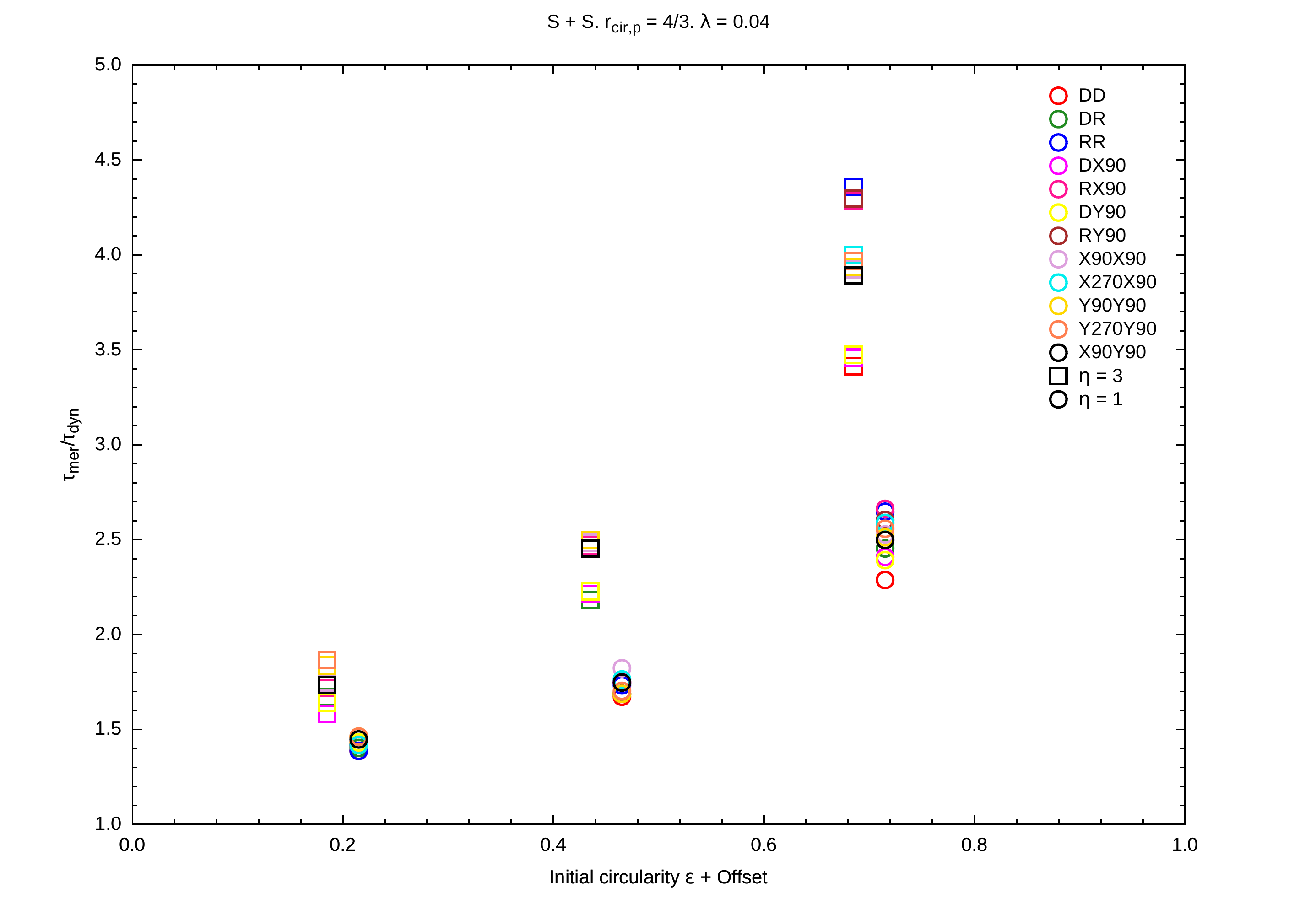}
	\caption{\small Same as Fig.~\ref{fig_5} but showing the effect of
		varying the mass ratio, $\eta$. Data points show results for
		mergers of spiral galaxies with the modal value of the orbital
		energy ($r_{\rm cir,p}=4/3$). Open squares are used for mergers
		with $\eta=3$, while open circles identify $\eta=1$
		mergers.}\label{fig_6}
\end{figure*}

\subsection{On the orbital parameters}

Figs.~\ref{fig_5} and \ref{fig_6} illustrate, respectively, the
effects of varying the initial orbital energy and mass ratio on the
merger time, in both cases as a function of the initial circularity of
the orbit. In all our simulations there is a clear trend that the full
length of the merger increases with these three parameters,
consistently with earlier studies of dynamical friction. The increase
in the value in either of the first two parameters delays the merger
more markedly as the circularity increases. Indeed, within the ranges
of values chosen for our simulations, we find that the affectations in
the location and scale of the merger time distributions induced by
varying $r_{\rm cir,p}$ and $\eta$ are similar, with the sole
difference that in the most radial collisions ($\epsilon=0.2$) the
changes in the orbital energy have barely any impact in $\tau_{\rm
  mer}$ ($\lesssim 5\%$). Also shown by these two plots is the fact
that the set of extremal initial relative orientations adopted for the
internal spins of the parent galactic haloes introduces a significant
scatter in $\tau_{\rm mer}$, which grows in line with the modulus of
$\lambda$. In the most favourable configurations among all the mergers
investigated (i.e.\ those with $r_{\rm cir,p}=2.0$, $\eta=3$,
$\lambda=0.06$) the spread can be up to $\pm 15\%$, equivalent to a
total time-span of about 3 Gyr. Nevertheless, it can be estimated, by
assuming that the spin plane of the main halo and the orbital angular
momentum plane are uncorrelated \citep{KB06}, that only $\lesssim 2\%$
of the pairs in a given configuration participate in the most extremal
deviations.

%Fig. 7
\begin{figure*}
	\centering
	\includegraphics[width=\linewidth,keepaspectratio,clip=true,angle=0]{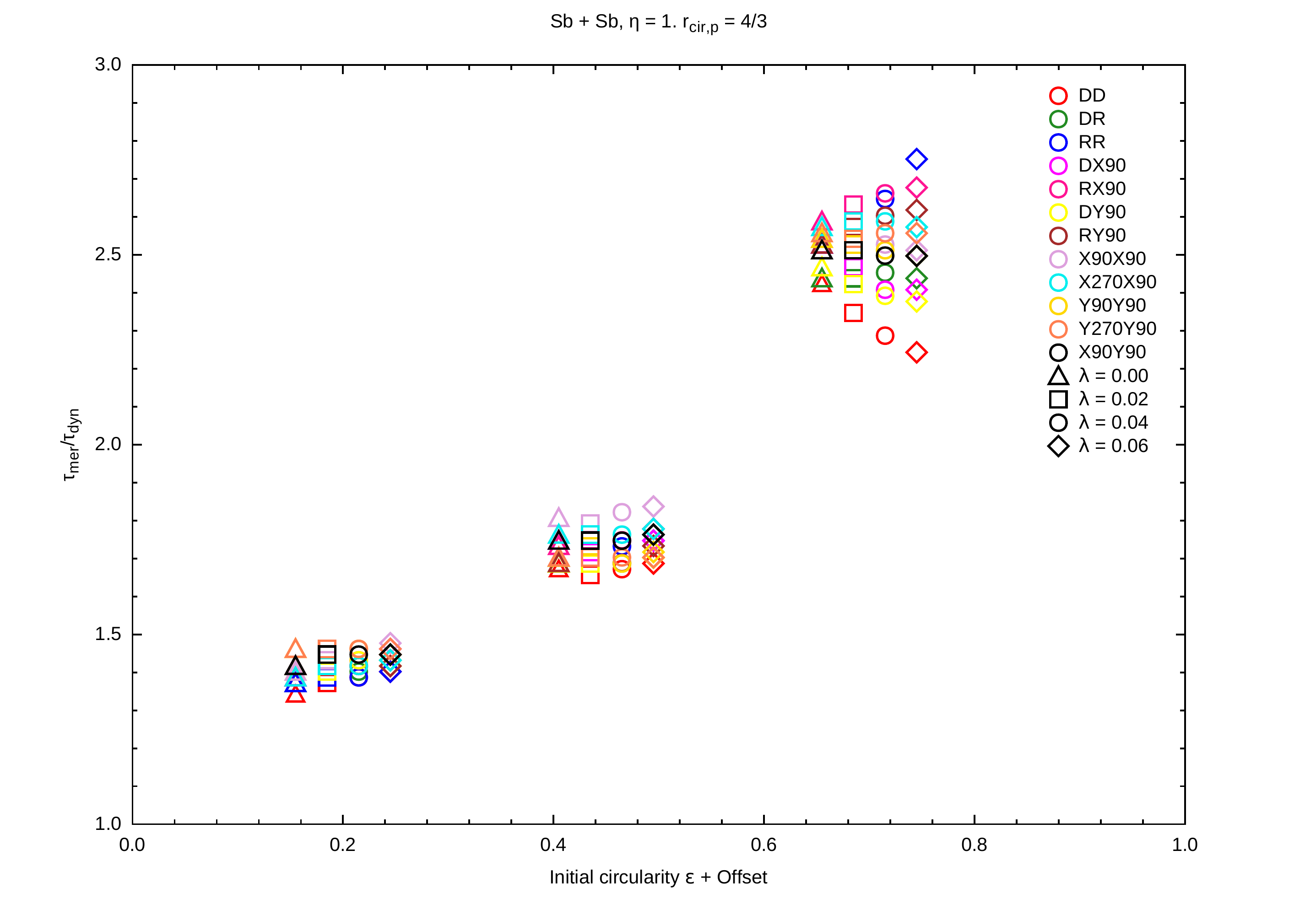}
	\caption{\small Same as Fig.~\ref{fig_5} but showing the effect of
		varying the halo spin parameter, $\lambda$. Data points show results
		for equal-mass mergers of Sb galaxies with the modal value of the
		reduced orbital energy ($r_{\rm cir,p}=4/3$). Open symbols identify
		different values of the spin parameter: $\lambda=0.00$ (triangles),
		$\lambda=0.02$ (squares), $\lambda=0.04$ (circles), and
		$\lambda=0.06$ (diamonds). Data points are shown with different
		offsets around the true values of $\epsilon$ for
		clarity. }\label{fig_7}
\end{figure*}

\subsection{On the internal spins of host and satellite}

The role of the progenitors' spin in the duration of mergers can be
better understood with the aid of Fig.~\ref{fig_7}. Since under normal
circumstances it is reasonable to expect that $J_{\rm orb} >> J_{\rm
  halo} >> J_{\rm stars}$\footnote{Even the fast-rotating stellar
  discs contribute modestly to the total spin of galaxies endowed with
  rotating massive DM haloes. E.g., in our primary late-type galaxies
  with $\lambda=0.04$ the stellar component provides less than $3.5\%$
  of the total internal angular momentum.}, a boost in $\epsilon$,
$\eta$, or $r_{\rm cir,p}$, which are all parameters having a direct
bearing on $J_{\rm orb}$, must easily bring about significant changes
on the length of the merger time. Besides, any increase (decrease) in
the value of the last two parameters while keeping all the other
constant automatically leads to an increase (decrease) on the relative
velocity of the collision, so on the amount of orbital energy, while
the efficiency at which this energy is transformed into internal
random motions decreases (increases) -- the force from dynamical
friction goes as $f_{\rm dyn}\propto M_{\rm p}\rho_{\rm s}/v^2$. The
strength of the dynamical friction prevailing on any merger is
therefore essentially controlled by $\epsilon$, $\eta$, and $r_{\rm
  cir,p}$. In contrast, changes in $\lambda$ only affect $J_{\rm
  halo}$, so this parameter is expected to have a lesser influence on
the merger time than the previous ones, which manifests as an increase
in the spread of $\tau_{\rm mer}$. The observed symmetry on the scale
of the deviations (see Fig.~\ref{fig_7}), relates to the symmetry of
the adopted spin configurations  -- remember that in our simulations both host and satellite are assumed to have the same spin parameter -- and the result, advanced in the
previous section, that in mergers involving mass ratios larger than
unity only the main progenitor's spin matters. This characteristic
also helps us to comprehend why the merger times of the simulations
with $\eta=3$ are not evenly allocated, but show a tendency to
aggregate into distinct peaks that become more evident as the orbital
circularity increases (see Fig.~\ref{fig_6}). Thus, although the
peculiar trimodal distribution obtained for the most circular
encounters is motivated in part by the extreme configurations adopted
for the initial relative orientations of the internal spins, its root
cause surfaces when one considers which configurations make up the
different groupings. In particular, we find that all set-ups in which both
internal spins are orthogonal to that of the orbital plane, i.e. those
identified in our simulations with the labels X90X90, X90X270, Y90Y90,
Y90Y270 and X90Y90, behave as collisions of spinless haloes, giving
rise to merger times which group around the central part of the time
distribution inferred for $\epsilon=0.7$. In three other
configurations, RR, RX90 and RY90, the spin of the host determines
that the collisions are basically retrograde, giving rise to the
longest merger times, all also with durations similar to each other,
while for the remaining four configurations, DD, DR, DX90 and DY90,
the direct rotation of the main halo enhances the effectiveness of the
encounters, producing the lower peak associated with the shortest
time-scales. These results are consistent with the outcome of the work
by \citet{Vil12} who studied the evolution of disc galaxies within the
global tidal field of the group environment, finding that discs on
retrograde orbits retain their internal structures and kinematics
longer, in comparison to prograde discs.

As regards the role of the satellites' spin, it seems to be restricted
to establish the degree of interaction of the stellar cores, which
determines the efficiency with which $J_{\rm stars}$ is drained. Thus,
if we compare, for instance, the configurations DY90 and DX90
belonging to the same mode, we see that the small difference in their
merger times arises because in the first case the relative
orientations of the galaxies favour a somewhat more intense contact
between them, which leads to narrower close approaches near
coalescence that increase slightly the speed of the orbital
decay. Interestingly, although the segregation of the time-scales
induced by $\lambda$ gradually develops from the first moments of the
collision, it has no effect on the total number of pericentric
passages experienced by the galaxies in the pre-coalescence stage,
which is the same (four) in all cases; nor we observe this number to
depend on the mass ratio of the progenitors or the ellipticity of the
orbit (compare Figs.~\ref{fig_2} and \ref{fig_3}).

\subsection{On the morphology of progenitors}

\begin{figure*}
\centering
\includegraphics[width=\linewidth,keepaspectratio,clip=true,angle=0]{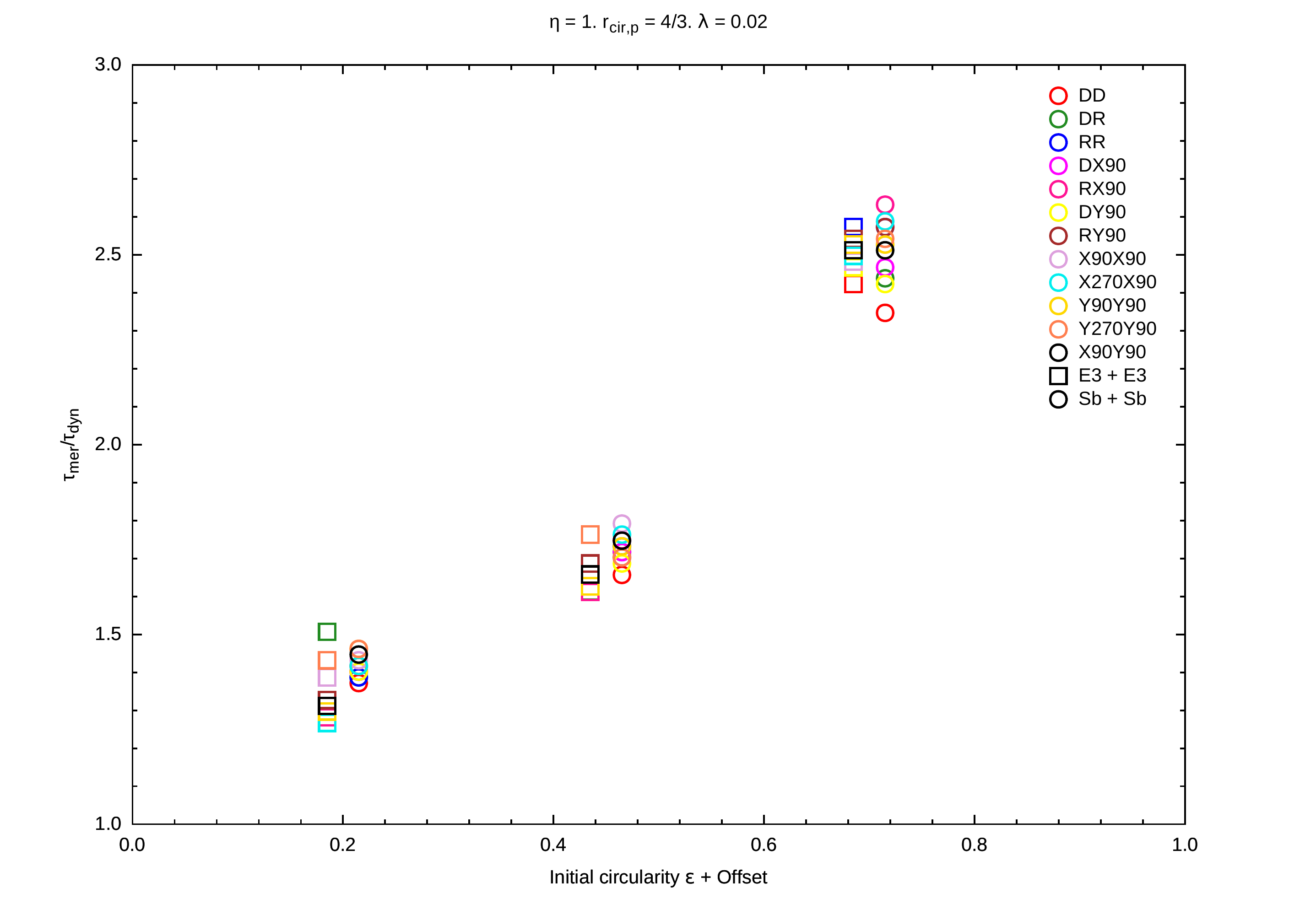}
\caption{\small Same as Fig.~\ref{fig_5} but showing the effect of
  varying the morphology of the central object. Data points show
  results for equal-mass mergers of galaxies with reduced orbital
  energy $r_{\rm cir,p}=4/3$ and halo spin parameter
  $\lambda=0.02$. Open squares are used for mergers of pairs of E3
  galaxies, while open circles identify Sb+Sb mergers.}\label{fig_8}
\end{figure*}

Lastly, we have also tested the effects of changing the internal
distributions of the baryonic component (stars) on merging
time-scales. As shown in Fig.~\ref{fig_8} mergers of pure stellar
spheroids are somewhat faster that those of identical characteristics
but involving disc galaxies, as expected from the fact that in the
first case the density in the innermost region of the galactic haloes
is larger. Nevertheless, since this difference only matters at late
stages in the merger, we observe a quite modest cropping of the merger
times: less than around $5\%$ and $8\%$, respectively, in highly and
moderately radial collisions, that favour the proximity of the centres
of host and satellite during the pre-coalescence phase, and virtually
zero for collisions along nearly circular orbits. Interestingly, in
mergers of elliptical galaxies the behaviour of the scatter associated
with the different internal spin orientations turns out to be a mirror
image of that of spirals, growing as the circularity diminishes.

%Fig. 8

\section{Major merger time-scales}\label{tau_mer_scales}

The most robust and accurate fitting formulas for the merger
time-scale obtained from simulations are revisions of
\citeauthor{LC93}'s \citeyear{LC93} analytical expression inspired by
the idealized description of the phenomenon of dynamical friction
given by \citet{Cha43}. These equations tie $\tau_{\rm mer}$ to the
orbital parameters that define the trajectory of any closed two-body
system and to the mass ratio between host and satellite, for which,
according to the dynamical friction theory, the merger time is
expected to show a rather strong positive correlation. The general
form of the models, which treat each factor contributing to the
dynamical friction separately, is given by
\citetext{c.f.~\citealt*{BKMQ08,Jia08,McCa12}}
\begin{equation}\label{tau_mer_formula}
\frac{\tau_{\rm mer}}{\tau_{\rm dyn}}=A\frac{\eta^B}{\ln (1 +
  \eta)}f(\epsilon;{\bf C})\left[\frac{r_{\rm cir}}{R_{\rm
      p}}\right]^D\;,
\end{equation}
where constants $B$ and $D$ parametrize the dependence of the merger
time-scale on the mass ratio $\eta$ and reduced orbital energy $r_{\rm
  cir,p}$, respectively, $\ln (1 + \eta)$ is a widely used
approximation to the Coulomb logarithm, $\ln \Lambda$, that works well
for major mergers \citep[e.g.][]{Jia08}\footnote{$\Lambda$ in the
  Coulomb logarithm $\ln \Lambda$ is defined as the ratio of the
  maximum and minimum impact parameters for which encounters between
  the satellite and the background sea of matter from the host can be
  considered effective; the ad-hoc approximation $\ln (1 + M_{\rm
    p}/M_{\rm s})$ is used in the literature to make the original
  \citeauthor{LC93}'s (\citeyear{LC93}) formula, which is valid for
  minor mergers, also applicable to major mergers.}, while
$f(\epsilon;{\bf C})$ is intended to represent the different
parametric functions that are used to encapsulate the dependence on
the orbital circularity $\epsilon$ (see Sections
\ref{tau_mer_comparison} and \ref{tau_mer_fits}). The factorization
also affects the dependence on the cosmology, which is exclusively
provided by the halo-independent dynamical time-scale:
\begin{equation}\label{tau_dyn}
\begin{split}
\tau_{\rm dyn} \equiv & \frac{R_{\rm halo}^{3/2}}{(GM_{\rm
    halo})^{1/2}} = \\ & 0.978 \left[\frac{\Omega_0\Delta_{\rm
      vir}(z)}{200}\right]^{-1/2}(1+z)^{-3/2}\,h^{-1}\;{\rm{Gyr}}\;,
\end{split}
\end{equation}
with $\Dvir$ the virial overdensity in the spherical collapse model \citetext{cf.\ \citealt{BN98}}:
\begin{equation}
\Dvir(z)\simeq\frac{18\pi^2+82[\Om (z)-1]-39[\Om (z)-1]^2}{\Om (z)}\;,
\end{equation}
and where the cosmological density parameter can be easily inferred from
\begin{equation}
\Om(z)=\frac{1}{1+(\Olo/\Omo)(1+z)^{-3}}\;,
\end{equation}
so for the adopted cosmology $\Dvir = 368$ at the present epoch.

To avoid ambiguities in the application of
equation~(\ref{tau_mer_formula}), the masses of host and satellite are
defined by convention as the virial masses of both objects right
before their haloes start to overlap. For the same reason, it is
preferable that the values of the remaining orbital parameters are
determined at that same instant, before significant interaction has
occurred, when the orbital energy is still conserved and galaxies
essentially behave as test particles moving along Keplerian orbits. By
using the asymptotic Keplerian values of the orbital parameters
instead of locally defined quantities, one makes the equation for the
merger time-scale independent of the details of the internal structure
of the merging haloes.

\subsection{Comparison of merger time-scales}\label{tau_mer_comparison}

We now compare the merger times obtained from our simulations to four
well-known empirical models of the merger time-scale existing in the
literature. These are models that obey
equation~(\ref{tau_mer_formula}) with different best-fit parameters
and circularity functions. One is the formula given by \citet{BKMQ08},
which we label hereafter as B08, which has $(A,B,D)=(0.216,1.3,1.0)$
and $f(\epsilon;{\bf C})={\rm{exp}(1.9\epsilon)}$; another is the
formula from \citet{McCa12}, labelled M12, which has
$(A,B,D)=(0.9,1.0,0.1)$ and $f(\epsilon;{\bf
  C})={\rm{exp}(0.6\epsilon)}$; finally, the last two expressions are
from \citet{Jia08}, one, labelled J08, which does not depend on the
energy of the satellite and is given by $(A,B,D)=(1.163,1.0,0.0)$ and
$f(\epsilon;{\bf C})=0.94\epsilon^{0.6}+0.6$, and other, labelled
J08$_{\rm e}$, which does, and is characterized by
$(A,B,D)=(0.163,1.0,0.5)$ and $f(\epsilon;{\bf
  C})=0.90\epsilon^{0.47}+0.6$.

%Fig. 9
\begin{figure*}
\centering
\includegraphics[width=0.9\linewidth,keepaspectratio,clip=true,angle=0]{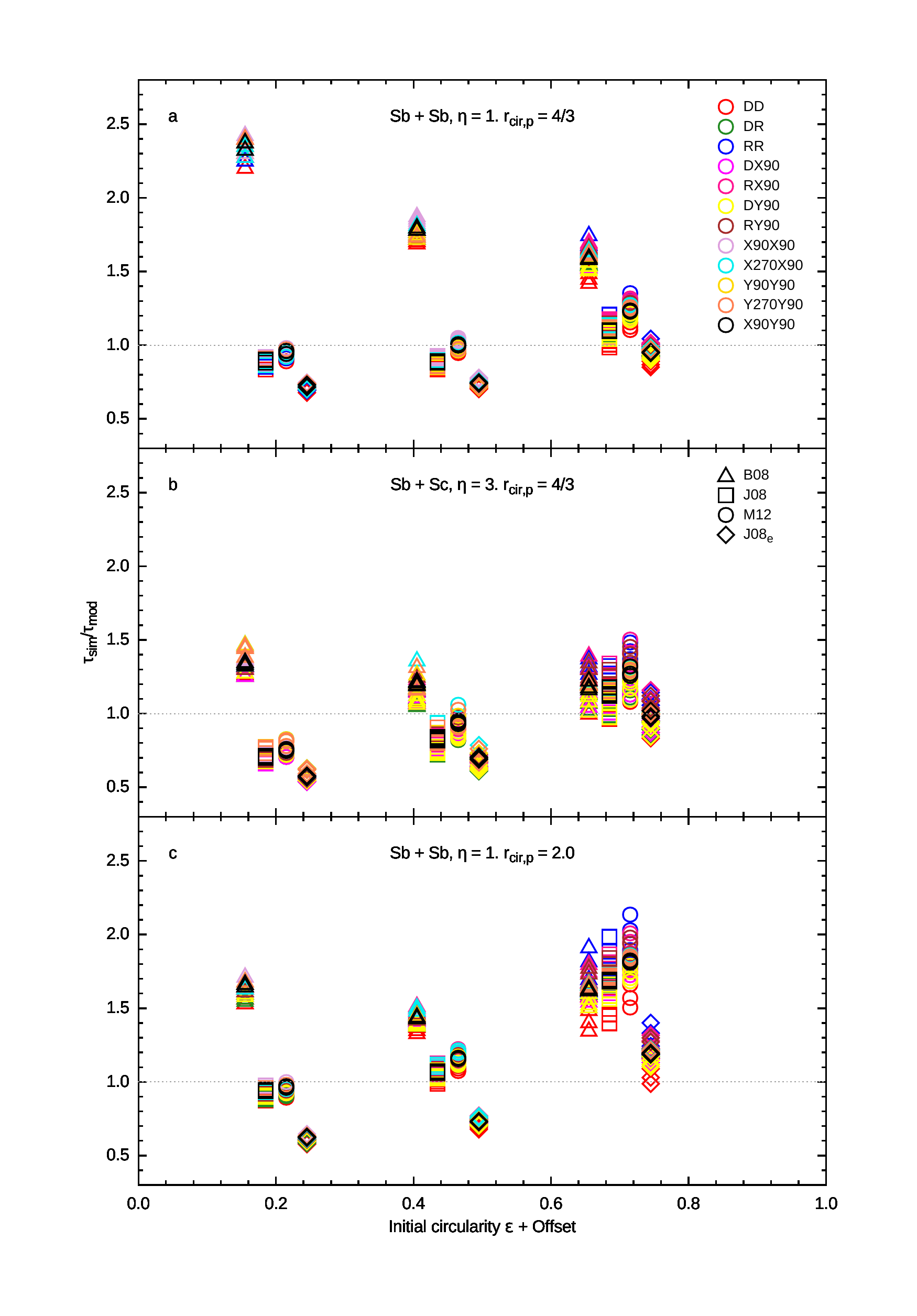}
\vspace*{-5mm}
\caption{\small Comparison between predictions from the B08, J08,
  J08$_{\rm e}$, and M12 models and our measurements for the case of
  merger simulations with a reduced orbital energy of $4/3$ and mass
  ratios 1:1 and 3:1 (panels a and b, respectively), and for
  equal-mass mergers with an initial energy of $2.0$ (panel c), as a
  function of the initial orbital circularity of the satellites. All
  ratios are shown with different offsets around the true values of
  $\epsilon$ for clarity.}
\label{fig_9}
\end{figure*}

To illustrate this comparison, we show in Fig.~\ref{fig_9} the ratios
between the merging times measured in our simulations, $\tau_{\rm
  sim}$, and the predictions of the B08, J08, J08$_{\rm e}$, and M12
models, $\tau_{\rm mod}$, from a subset of our experiments involving
disc mergers with reduced orbital energy $r_{\rm cir,p}=4/3$ and mass
ratios $\eta=1$ and $3$ (top and middle panels, respectively), and
with $r_{\rm cir,p}=2.0$ and $\eta=1$ (bottom panel). The analysis of
the predicted merger time-scales as a function of the initial orbit of
the galaxies shows quite uneven results, with a reduced level of
agreement between the B08 model and those with a cosmological basis,
especially for head-on encounters. The largest differences between
the predictions and our experiments are also found with respect to the
B08 model, which significantly underpredicts major merger times. For
mergers with the most likely orbital energy ($r_{\rm cir,p}=4/3$),
this model underestimates the time scale of pairwise radial encounters
by a factor of $\sim 2.3$. The comparison to B08 produces much tighter
agreement for 3:1 mergers, where this model exhibits fractional
deviations of comparable strength ($\lesssim 30\%$) but different sign
to the other functional forms. The relatively important disagreement
between our measurements and the predictions from the B08 formula is
somewhat surprising since the latter was derived from simulations of
isolated mergers of the same type as those presented in this work. We
note, however, that although \citet{BKMQ08} experiments and ours cover
a similar range of initial circularities, the ranges of both the
initial orbital energy and mass ratio of the mergers are quite
different \citetext{in particular, \citeauthor{BKMQ08} dealt with
  minor mergers having $\eta\leq 0.3$ and reduced energies less than
  one}. This suggests that at least part of the discrepancy may be
attributable to the fact that our simulations fall beyond the range of
validity of their approximation. Another factor that may contribute to
the observed mismatch are differences in the definition of the merger
endpoint. B08 used a $5\%$-threshold in the specific orbital spin of
their dark haloes, while we are able to track the coalescence of the
baryonic cores of the galaxies. As discussed in
Section~\ref{tau_merge} (see also Fig.~\ref{fig_4}) the use of the
specific angular momentum to monitor the final act of the orbital
decay systematically underestimates $\tau_{\rm mer}$ in an amount that
increases with increasing orbital eccentricity. However, 
any possible bias attributable to the metric appears to fall short of
explaining the significant deviation detected.

In contrast, the comparison to the two J08 models and the M12 one, all
derived from mergers simulated in a cosmological context, shows a
better agreement. Independently of the initial value of the orbital
energy and mass ratio, the J08 and M12 formulas tend to underestimate
(overestimate) $\tau_{\rm mer}$ for galaxies with higher (lower)
initial circularity, with both the J08 and M12 models performing in
much the same way overall. On the other hand, the J08$_{\rm e}$
functional form appears to be fine tuned to accurately predict
time-scales for mergers in highly circular orbits, but the predictions
for moderate and nearly radial encounters are somewhat worse than
those from the other two formulas calibrated with cosmological
experiments. Although the values of the deviations are seemingly not
as problematic as those resulting from the comparison with the B08
model, their sign and the positive correlation that the ratio
$\tau_{\rm sim}/\tau_{\rm mod}$ shows with the circularity are. Thus,
for example, the tendency of the M12 model to overestimate $\tau_{\rm
  mer}$ in nearly radial encounters, especially noticeable when
$r_{\rm cir,p}=4/3$, is inconsistent with the utilization by
\citet{McCa12} of a metric based on the minimization of the orbital
angular momentum. Things are much the same with J08, where shorter
merger times may be expected from the overcooling problem that
affected their simulations \citep*{JJL10}. In the same vein, an
explanation based on the different context in which the simulations
have been performed, isolated versus cosmological, is unlikely to work
either. If this were the case, one would expect the time-scales
inferred from our simulations to show a progressive tendency to be
biased low with respect to the cosmologically founded formulas with
increasing $\epsilon$, contrarily to what is observed -- arguably,
environmental effects should progressively delay mergers as time goes
by. The same goes for other factors, such as the possible differences in the internal structure of the halos that would result from the fact that we have ignored the response of dark matter to the condensation of baryons, since the adiabatic contraction of our dark halos would have increased its central densities and, therefore, decreased $\tau_{\rm mer}$, thus further increasing the inconsistency. Nor does it seem likely that the discrepancies can be attributed to large differences in the respective parameter spaces. In contrast to
the B08 model, the values of orbital circularities and mass ratios
adopted in our simulations fall squarely within the range of validity
of the cosmological expressions, whilst the reduced orbital energies
show a partial overlap only with respect to the study by
\citet{Jia08}, which deals with simulations limited to values of this
parameter below $1.5$.

%Fig. 10
\begin{figure*}
	\centering
	\includegraphics[width=\linewidth,keepaspectratio,clip=true,angle=0]{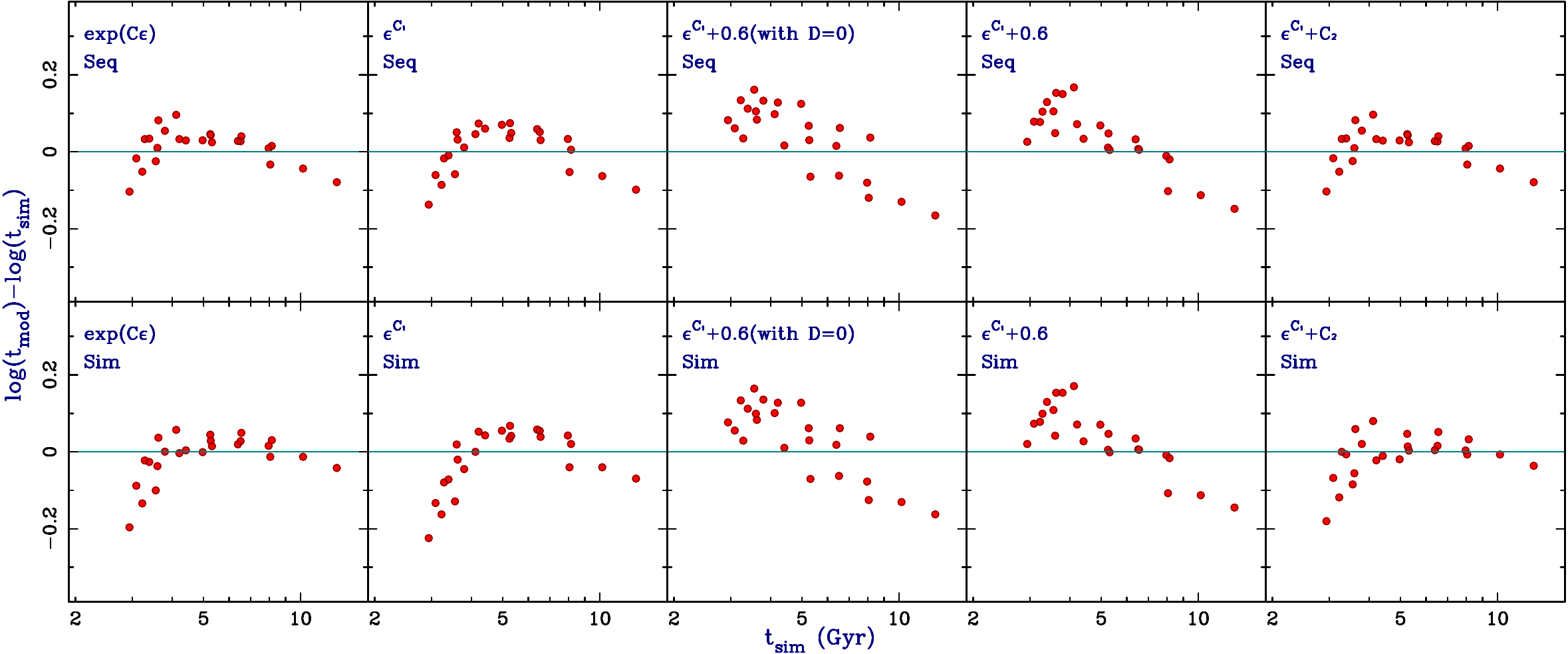}
	%\vspace*{-5mm}
	\caption{\small Residuals of the comparison between the merger time-scales measured 	         in our simulations and the different predictions arising from the best
		fits of the models listed in Table \ref{model_fits}. The
		plots are organized in two rows, corresponding to the two fitting
		procedures considered (top: (Seq)uential; bottom: (Sim)ultaneous), and
		five columns showing the different functionals for $f(\epsilon;{\bf
			C})$ adopted in Table~\ref{model_fits}. The log-log scale adopted makes it easier to appreciate the magnitudes and trends of the deviations.}
	\label{fig_10}
\end{figure*}

Finally, we have also discarded the possibility that the largest
discrepancies between our simulations and the different predictions
arise from any of the arguments put forward in the former comparison
work by \citet{Vil13}. These authors have conducted experiments of
isolated mergers between a single galaxy and a much larger main halo
at different redshits finding that prescriptions based on
equation~(\ref{tau_mer_formula}) significantly underestimate long
merger time-scales. \citeauthor{Vil13} attribute the mismatching to
the neglect of the evolution of halo concentration in controlled
experiments and to the possible undersampling of long-term mergers in
cosmological simulations. In both cases, they are able to partially
compensate the observed offsets by including in
equation~(\ref{tau_mer_formula}) an extra term of the form $(1+z)^E$,
with $E\sim 0.45$--$0.60$. We note, however, that this is a correction
designed for very long-term mergers ($\gtrsim 8\tau_{\rm dyn}$), so it
is not applicable to the moderately-long mergers investigated in this
paper.

%Table 3
\begin{table*}
%\begin{minipage}{140mm}
%\scriptsize
\centering
%\rotate
\caption{Best-fit parameter values and fractional rms error of the residuals from fits of eq.~(\ref{tau_mer_formula}) to our major merger simulations.}
\label{model_fits}
%\begin{threeparttable}
%\tablecolumns{8}
%\tablewidth{0pt}
%\tablenum{1}
\begin{tabular}{llccccccc}
\hline\hline

%\tabletypesize{\scriptsize}
Fitting method & \multicolumn{1}{l}{$f(\epsilon;{\bf C})$} & $A$ & $B$ & $C$ &  $C_1$ & $C_2$ & $D$ & RMSE \\
\hline
Sequential & $\exp(C\epsilon)$ & .461 & 0.97 & 1.84 & - & - & 0.61 & 0.11  \\
 & $\epsilon^{C_1}$ & 2.20 & 0.97 & - & 0.78 & - & 0.56 & 0.14 \\
 & $\epsilon^{C_1}+C_2$ & 1.84 & 0.97& - & 1.73 & $\:\,0.60^i$ & $\,0.00^i$  & 0.24 \\
 & $\epsilon^{C_1}+C_2$  & 1.44 & 0.97 & - & 1.73 & $\:\,0.60^i$ & 0.42 & 0.22 \\
 & $\epsilon^{C_1}+C_2$ & 2.15 & 0.97 & - & 1.81 & $\: 0.26$ & 0.61 & 0.11 \\
 \hline
Simultaneous & $\exp(C\epsilon)$ & .330 & 1.00 & 2.26 & - & - & 0.74 & 0.13 \\
 & $\epsilon^{C_1}$ & 2.26 & 1.00 & - & 0.94 & - & 0.64  & 0.16 \\
 & $\epsilon^{C_1}+C_2$ & 1.82 & 0.99 & - & 1.73 & $\:\,0.60^i$ & $\,0.00^i$  & 0.24 \\
 & $\epsilon^{C_1}+C_2$  & 1.42 & 0.99 & - & 1.76 & $\:\,0.60^i$ & 0.42 & 0.22 \\
 & $\epsilon^{C_1}+C_2$ & 2.66 & 1.00 & - & 2.40 & $\: 0.18$ & 0.76  & 0.12 \\
\hline\hline
\citet{LC93}$^{ii}$ & $\epsilon^{C_1}$ & $.581$ & 1.0 & - & 0.78 & - & 2.0  & N/A \\
\citet{BKMQ08} & $\exp(C\epsilon)$ & $.216$ & $\;\; 1.3^{iv}$ & 1.9 & - & - & 1.0 & N/A \\
\citet{Jia08} & $0.94^{iii}\epsilon^{C_1}+C_2$  & $\:\:1.16^i$ & $\; 1.0^i$ & - & 0.60 & $\:\:0.60^i$ & $\:\:0.0^i$  & N/A \\
& $0.90^{iii}\epsilon^{C_1}+C_2$ & $\:\:1.16^i$ & $\,\,1.0^i$ & - & 0.47 & $\:\:0.60^i$ & $\,0.5$  & N/A \\
\citet{McCa12} & $\exp(C\epsilon)$ & $\:\:0.9$ & 1.0 & 0.6 & - & - & $\,0.1$ & N/A \\
\hline
\end{tabular}

\tablefoot{$^i$Fixed a priori. $^{ii}$Coefficients determined
  analytically. $^{iii}$Fitted. $^{iv}$According to \citet{BKMQ08} the
  merger times from their most major merger simulations are somewhat
  better fit by $B=1$ than by $B=1.3$. N/A stands for non-available.}
%\end{minipage}
\end{table*}

%Two-Steps & $exp(c \, \varepsilon)$ & 1.29$/\tau_{dyn}$ & 1.00 & 1.85 & - & - & 0.11\\
 %& $c_{1}+\varepsilon^{c_{2}}$ & 6.06$/\tau_{dyn}$ & 1.00 & - & 0.26 & 1.82 & 0.11  \\
 %& $\varepsilon^{c_{2}}$ & 6.00$/\tau_{dyn}$ & 1.00 & - & - & 0.78 & 0.10  \\
 %& $0.6+\varepsilon^{c_{2}}$ & 3.65$/\tau_{dyn}$ & 0.99 & - & 0.6 & 1.73 & 0.08 \\

\subsection{Adapting existing prescriptions to major mergers}\label{tau_mer_fits}

Deriving a new, more accurate prescription to estimate major merger
time-scales is beyond the scope of this work. Instead, we have chosen
to recalculate the coefficients of the general formula
(\ref{tau_mer_formula}) provided above -- using both exponential and
power-law functions for the orbital angular momentum factor as
proposed in the literature -- by fitting it to our results, so we
produce expressions suitable for major mergers.

To perform this task, the results from our simulations have been
reduced to a three-dimensional grid in the space
$(\eta,\epsilon,r_{\rm cir,p})$ by taking the median values over all
relative orientations and moduli of the internal galactic spins,
resulting in a total of 27 data points. Just like in earlier studies,
we give the same weight to each one of these average values.

We have accounted not just for the different proposals made in the
literature for $f(\epsilon;{\bf C})$, but also for the different
strategies used in the fits. On the one hand, some authors do a
maximum likelihood estimate of all variables simultaneously, holding
fixed or not the values of some of parameters
\citep[e.g.][]{BKMQ08}. On the other hand, \citet{Jia08} and
\citet{McCa12} apply a sequential procedure, fitting the free
parameters of the prescriptions associated with each one of the
variables ordered from more to less impact on the merger time: first
$\eta$, then $\epsilon$, and finally $r_{\rm cir,p}$. If, as expected,
the magnitudes involved in the calculation of $\tau_{\rm mer}$ are
independent, the two methods should give very similar results,
otherwise the order of the fits matters\footnote{Some testing with our
  data has shown that initiating the sequence of the fits using either
  $\eta$ or $\epsilon$ introduces no significant variations in the
  parameter estimates.}. The combination of expressions for the
circularity and fitting strategies allows a total of ten different
fitting variants for the parameters of equation
(\ref{tau_mer_formula}) that we list in Table \ref{model_fits}. For
ease of comparison, this Table also provides the values attributed in
the original works to these parameters on which our procedure is
based.

We now attempt to gain some insight by examining the similarities and
differences among the outcomes from our fits. In the first place,
Table \ref{model_fits} shows that the mass dependence for $\tau_{\rm
  mer}$ adopted in equation~(\ref{tau_mer_formula}) with $B\approx 1$
provides a good match for the whole range of mass ratios, regardless
of whether major mergers are performed in either an isolated or a
cosmological context. In contrast, both the circularity and (reduced)
energy dependencies exhibit disagreements that are hard to
dismiss. Thus, when using an exponential form for the circularity, the
fits to our numerical results give exponents not too different from
the $1.9$ value found by \citet{BKMQ08}, but far from the $0.6$
adopted in the M12 model. On the other hand, fits to $\tau_{\rm mer}$
using a pure power-law expression for the circularity yield exponents
identical o somewhat higher than the value of $0.78$ analytically
derived by \cite{LC93}, while those based on the linear power-law
introduced by \citet{Jia08} lead to substantially stronger
dependencies, even when the zero offset is forced to be $0.6$ as these
authors did to ensure that their formula gives $\tau_{\rm
  mer}\approx\tau_{\rm dyn}$ for equal-mass mergers\footnote{Our
  simulations show that this is not the case even for highly radial
  orbits (see, for instance, our Fig.~\ref{fig_4}).}. Regarding the
pure power-law form also adopted for the dependence on the orbital
energy, we find that the values of the exponent in the range
$0.43$--$0.77$ provide the best match to our major-merger data. This
is in contrast with the canonical value of $D=2$ derived analytically
\citep[e.g.][]{BT87,Taf03}, but in line with previous numerical
simulations, which find exponents equal to or less than one. The
agreement is indeed rather good with respect to the J08$_{\rm e}$
model, for which $D=0.5$, and somewhat less satisfactory for the B08
model, for which $D=1.0$, while the dependence obtained by the M12
model $D=0.1$ seems excessively weak according to our results. On the
other hand, the J08 model does not include an explicit orbital energy
dependence, so $D=0$, but this is done to keep their fitting formula
simple.

In Fig.~\ref{fig_10} we plot the residuals resulting
from the comparison between the merger time-scales computed according to
the different variants of equation (\ref{tau_mer_formula}) listed in
Table \ref{model_fits} and the average merging times measured from
our simulations. As the last column of Table
\ref{model_fits}) demonstrates, the two fitting
procedures produce similarly accurate results, with a slender and
insignificant advantage for the sequential fitting 
regarding the size of the standard deviation of the residuals or RMSE
-- in relative terms it ranges from around $11\%$ to $24\%$. The good
performance of the fits does not hide, however, the presence of
a clear pattern in the point clouds of the Figure which indicates that the models are
not entirely accurate. This may help to explain, for instance, the substantial 
variations in the value of the power $D$ of the $r_{\rm cir,p}$
function reported in the literature (see the corresponding column in Table
\ref{model_fits}). Furthermore, regarding the different alternative
forms adopted for $f(\epsilon;{\bf C})$, Table~\ref{model_fits} also shows that
the approximating formulas with an exponential, $\exp(C\epsilon)$, and 
a full linear power-law dependence, $\epsilon^{C_1}+C_{2}$, provide
the tightest match to our data, with nearly identical values of the RMSE. Since the former expression only has one free parameter, it can be considered the best choice of 
all the investigated ones to describe the circularity dependence of the time-scale of major mergers from equation (\ref{tau_mer_formula}). On the other hand, the pure power-law fit, $\epsilon^{C_1}$, shares with the two former expressions a similar but somewhat more extended non-linear distribution of residuals, while the two power-law models with a zero offset of $0.6$ offer the worst performance in terms of the RMSE. However, the residuals of these latter models are roughly linearly distributed, so they should be easier to minimise when it comes to improving the prediction of $\tau_{\rm mer}$.

\section{Discussion and conclusions}\label{conclusionsI}

This work gives an in-depth examination of the orbital decay time for
major mergers of galaxies, a centerpiece of cosmic evolution. This has
been accomplished with the aid of nearly 600 high-resolution $N$-body
simulations of collisions of isolated pairs of similarly massive
galaxies on bound orbits. Our dissipationless experiments follow the
merging of live models of late- and early-type objects, composed of a
rotating NFW dark matter halo and a central stellar core, with mass
ratios ranging from 1:1 to 3:1, and with orbital parameters and global
properties broadly consistent with the predictions of the standard
$\Lambda$CDM cosmology. These simulations have been used to perform a
detailed analysis of the merger times driven by dynamical friction,
which ranges from its definition and the factors that determine its
length, to the comparison of our inferred values for $ \tau_{\rm mer}$
with predictions drawn from approximating formulas existing in the
literature. In spite of the fact that we are dealing with gas-free
experiments, we expect our conclusions to apply to both dry and (moderately) wet
mergers, as the baryonic physics is known not to have a strong impact
on the merger time-scale \citep[e.g.][]{JJL10,Col14}.

Regarding the first two aspects commented above, the main and newest
results of the present work are:
\begin{enumerate} %[wide, labelwidth=!,topsep=0pt]
\itemsep0em
\item We have identified the need of a unified criterion to qualify
  binary galaxy mergers. This is not a minor issue, as apparently
  small differences in the definition of the virial radius of host and
  satellite, in the input values of the orbital parameters, or in the
  start and end points of the merger, can lead to relatively large
  divergences in the determination of $\tau_{\rm mer}$. In
  particular, the strategy commonly used in numerical experiments of
  defining the time of coalescence of the galaxies as the moment at
  which the specific orbital angular momentum of the system reaches
  the $5\%$ of its initial value, $\tau_{\rm j05}$, has been shown to
  systematically underestimate the merger time-scale, resulting in
  fractional deviations that, in major mergers, can be as large as
  $60$ per cent for nearly radial collisions. Accordingly, to
  determine the time of coalescence in this kind of mergers, we
  recommend using instead the instant in which the time derivative of
  the logarithm of the product of the relative distance, $r$, and
  speed, $v$, of the progenitors' cores reaches its absolute
  maximum. For ease of identification, it may be necessary to apply a
  low-pass filter to the function $rv(t)$.
\item In major mergers, the internal spin of the parent galaxies may
  lead to variations of $\tau_{\rm mer}$ above $15$ per cent in highly
  circular collisions. In our simulations this corresponds to
  differences of up to $\sim 3$ Gyr in the duration of mergers. This
  implies that in controlled experiments the distribution of the
  initial relative orientations of the internal spins can potentially
  bias the outcome. In the real universe, however, the largest
  deviations in $\tau_{\rm mer}$ are expected to affect only a small
  number of pairs ($< 2\%$) within a given orbital configuration.
\item Except in nearly radial encounters, the initial orientation of
  the internal spin of the host regulates the strength of the
  dynamical friction prevailing on any merger. In contrast, the
  duration of major mergers -- much as with minor collisions -- is
  largely independent of the initial orientation of the internal spin
  of the satellite. Neither this spin, nor that of the main halo, nor
  the mass ratio of the progenitors, nor the circularity of its orbit,
  influence the number of pericentric passages in the pre-coalescence
  stage, which is determined exclusively by the orbital energy of the
  merger.
\item In accordance with expectations, binary mergers of pure stellar
  spheroids are somewhat faster that those of identical
  characteristics but involving discs. In our major mergers the
  differences are, however, quite modest, as we measure fractional
  deviations in $\tau_{\rm mer}$ that are $\lesssim 8\%$ in encounters
  that take place following very or fairly elongated orbits and
  negligible for those configurations in which the two merger partners
  spend most of their time at large separations. This is in sharp
  contrast with the results of the study of tidal stripping efficiency
  on minor mergers by \citet{CMK13}, who conclude that the morphology
  of small satellites has a very strong effect on the efficiency of
  stellar stripping.
\end{enumerate}

On the other hand, the comparison of the merger times obtained from
our simulations to the predictions arising from some of the
better-known empirically derived formulas existing in the literature
has revealed unexpected important discordances. We find that the B08
formula, which like this work relies on isolated mergers,
systematically underpredicts $\tau_{\rm mer}$, concentrating the
biggest variations (of more than a factor of two) on the most radial
encounters. This behaviour is qualitatively consistent with the use of
$\tau_{\rm j05}$ by these authors to measure merger time and the fact
that their simulations are biased against major
encounters\footnote{Mass losses via tidal stripping and gravitational
  shocking against merger companions are relatively more important for
  more massive galaxies \citep{Vil12}, so our pairs should merge more
  slowly than minor-merger-based formulas predict.}. Quantitatively,
however, the poor agreement found between the magnitude of the
predictions and our data, especially for the equal-mass case, suggests
that other factors, beyond differences in the metric or in the ranges
of the orbital parameter values used in both simulations, could be
contributing to the observed discrepancy. For the expressions with a
cosmological basis the situation is quite the opposite. In this case,
our numerical values are, as a whole, relatively well described by J08
and M12 approximating formulas (for some parameter combinations the
match is actually very good), but the signs and trends of the
differences are eye-catching. In particular, we have found that a
great deal of individual predictions overestimate our observed merger
times. This is contrary to expectations since some specific features
of cosmological simulations suggest that they should also show a
tendency to anticipate the coalescence of the satellite and the
central galaxy. Furthermore, and perhaps more importantly, in all the
cases investigated the differences between our merger time-scales and
the cosmological predictions show a trend with the orbital circularity
that is at odds with what might be expected from possible
environmental effects. Again, differences in the metric, the physics
and/or the context of the simulations do not appear sufficient to
explain the characteristics of the biases we observe.

Finally, the adaptation to our data of the coefficients of well-known
time-scale models -- generalised through the use of five alternative
expressions for the dependence of circularity and two different
fitting strategies --, has allowed us to demonstrate that the form
adopted in equation~(\ref{tau_mer_formula}) for the relation between
$\tau_{\rm mer}$ and the mass ratio works well for all kind of mergers
and contexts provided that the exponent $B\approx 1$. In contrast, we
find a relatively low degree of agreement among the values of the
coefficients of the expressions used to model the dependence of the
merger time on the circularity and, especially, the initial orbital
energy, that is not exclusive of major mergers. Such a lack of
consensus that stems, seemingly, from a description of the data that
is not entirely suited, has prevented us from drawing any firm
conclusion about which relationship would be the most adequate. These
findings suggest that the analytical prescriptions traditionally
adopted to describe the dependence of $\tau_{\rm mer}$ on the orbital
parameters, or at least some parts of them, are open to improvement.

Therefore, our concluding remark would be that there is still
considerable ground to be covered in the modelling of a problem as
complex as that of dynamical friction. To begin with, we have shown
the need to reach a general consensus on the quantification of the
fiducial starting and end points of mergers, as well as on the
definitions of the parameters governing the merging time, which have
to be clear and unambiguous. In addition, future investigations of
dynamical friction (aka merging) must take steps to perform a complete
identification of the physical mechanisms governing the time-scales of
the orbital decay. Only in this way will we be able to extend the
first principles contained in the original \citeauthor{Cha43}'s theory
to live objects assembling in cosmologically relevant simulations, to
replace its simplifying assumptions by realistic considerations and,
ultimately, to derive a single formula for the merger time-scale that
is accurate and valid across the entire range of mass ratios and
orbital parameters of the parent galaxies.

\section*{Acknowledgements}

The authors acknowledge financial support from the Spanish AEI and
European FEDER funds through the research project AYA2016-76682-C3 as
well as from the Program for Promotion of High-Level Scientific and
Technical Research of Spain under contract AYA2013-40609-P.

%%%%%%%%%%%%%%%%%%%%%%%%%%%%%%%%%%%%%%%%%%%%%%%%%%



%\bibliographystyle{mnras}

%\singlespace

%%%%%%%%%%%%%%%%% APPENDICES %%%%%%%%%%%%%%%%%%%%

\appendix

\section{Galactic halo rotation}\label{halorot}

Inspired by \citet{SW99}, we model the mean streaming velocity of the dark haloes,
$\overline{v_\phi}$, -- which in a static, axisymmetric system can be freely chosen to set its rotation --, as some fixed fraction, $f_{\rm h}$, of the local value of the radial velocity
dispersion used to generate the DM particle speeds, i.e.
\begin{equation}\label{streaming}
\overline{v_\phi}=f_{\rm h}\sigma_{\rm r}=f_{\rm h}\alpha v_{\rm c}\;,
\end{equation}
where
\begin{equation}\label{sigmar}
\sigma^2_{\rm r}(r)=\frac{1}{\rho(r)}\int_r^\infty \rho(r)\frac{GM(r)}{r^2}{\rm{d}}r\;,
\end{equation}
for a DM halo with an isotropic velocity tensor, and
$\alpha\equiv\langle\sigma_{\rm r}/v_{\rm c}\rangle_{r\leq R}$ is a
factor that we introduce to explicitly take into account the
expectation value of the ratio between the scale of DM particle
speeds, $\sigma_{\rm r}$, and the local azimuthal circular velocity,
$v_{\rm c}$, inside the halo virial radius. This factor is
model-dependent and must be determined empirically.

With $\alpha$ fixed, conservation of the specific angular momentum
during galaxy formation -- a reasonable assumption for late-type
galaxies but certainly not for merger-made early-type objects -- leads
to the conservation of $f_{\rm h}$ too, which can therefore be
computed from the total angular momentum $J$ of the initial halo. To
infer the latter, one must integrate the angular momenta of
infinitesimally thick spherical shells, ${\rm d} J=(2/3)rv_{{\rm
    c}}{\rm d}M$ up to the virial radius, which from the azimuthal
streaming model given by equation~(\ref{streaming}), a NFW halo mass
density profile, and the fact that, for a spherical mass distribution,
the maximum azimuthal circular speed is achieved at the equator, so
\begin{equation}
\left. v^2_{\rm{c,max}}\equiv R\frac{\partial\Phi(R,z)}{\partial R}\right|_{z=0}=\frac{GM(r)}{r}\;,
\end{equation}
gives:
\begin{equation}\label{angmom}
\begin{split}
J= & \frac{2}{3}4\pi f_{\rm h}\alpha\int_0^R r^2\rho(r)v_{\rm c}(r)r{\rm d}r=  \\ & \frac{2}{3}f_{\rm h}\alpha\left\{\frac{{GM}^3\rs}{[\ln(1+c)-c/(1+c)]^3}\right\}^{1/2}\gc\;,
\end{split}
\end{equation}
where $R$ and $M$ are the total (virial) radius and mass of the galactic halo, $\rs$ is the characteristic radius of the NFW density profile and $\gc$ is the integral
\begin{equation}\label{gc}
\gc\equiv\int_0^c \left[\ln(1+x)-\frac{x}{1+x}\right]^{1/2}\frac{x^{3/2}}{(1+x)^2}\;{\rm{d}}x\;,
\end{equation}
with $c\equiv R/\rs$ the usual definition of the concentration parameter. Combining this result with the definition of the dimensionless spin parameter $\lambda$ \citetext{c.f.~\citealt{Pee69}},
\begin{equation}\label{spin_param}
\lambda\equiv\frac{J\mid E\mid^{1/2}}{GM^{5/2}}\;,
\end{equation}
and taking into account that the total energy $E$ of a galactic halo
can be inferred from the virial theorem by assuming that all DM
particles move on circular orbits around the centre\footnote{Strictly,
  isolated haloes satisfy the scalar virial theorem in the limit
  $r\rightarrow\infty$. At the virial radius, objects with NFW
  profiles are only near dynamical equilibrium, with those of smaller
  mass and lower velocity anisotropy being closer \citep{LM01}.}, so
\begin{equation}\label{Ekin}
\mid E\mid=E_{\rm{kin}}=2\pi\int_0^R r^2\rho (r)v_{\rm c}^2(r){\rm{d}}r=\frac{GM^2}{2R}\fc\;,
\end{equation}
with
\begin{equation}\label{fc}
\fc=\frac{c}{2}\frac{1-1/(1+c)^2-2\ln(1+c)/(1+c)}{[\ln(1+c)-c/(1+c)]^2}
\end{equation}
for a NFW halo. In the present work those definitions of parameters that depend on redshift (e.g.~eq.~(\ref{M-c})) have been calculated at $z = 0$.

\section{Solving the two-body problem for extended objects}\label{one-body eqs}

%Fig. B1
\begin{figure*}
\vspace{-2cm}
	\centering
	\includegraphics[angle=-90,width=\linewidth,keepaspectratio,clip=true]{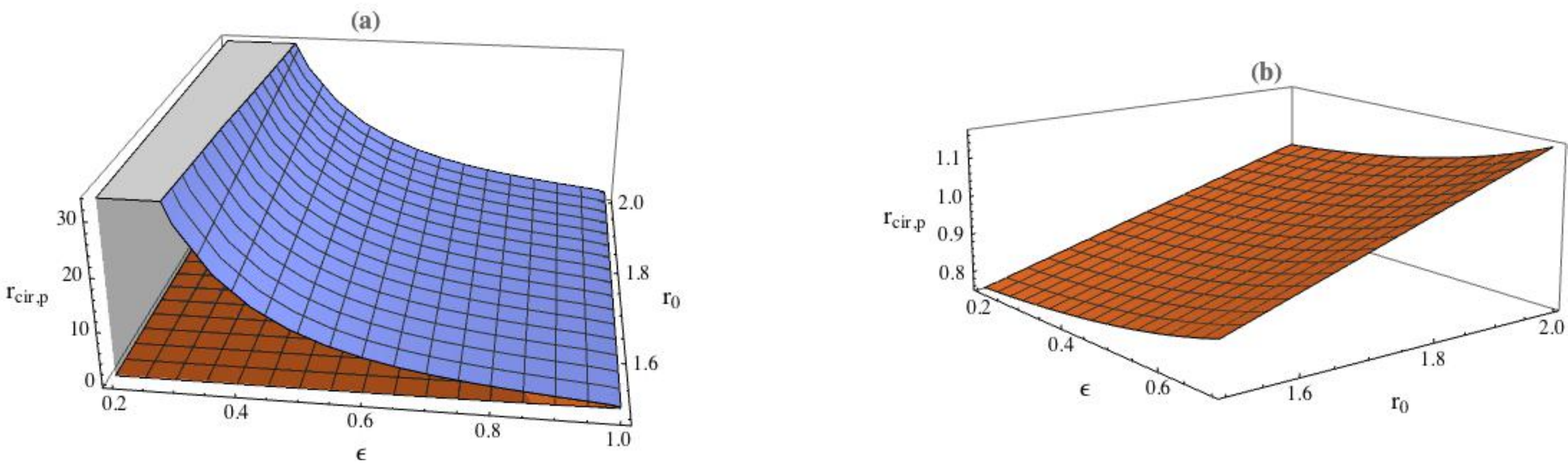}
	\vspace*{-30mm}
	\caption{\small (a) Portion of the domain of definition of the
		initial orbital parameters $\epsilon$, $r_0$, and $r_{\rm cir,p}$
		involved in the Keplerian two-body problem (shaded in gray). The
		blue and brown surfaces depict limiting cases. (b) A closer view of
		the lower boundary of this domain (brown surface).  It allows for a
		greater appreciation of the values of the initial orbital energy,
		$r_{\rm cir,p}$, below which the initial relative velocity, $v_0$,
		is not defined.}\label{fig_B1}
\end{figure*}

We summarize here the most important equations involved in the
standard approach followed to solve the motion of two extended 
objects interacting via a gravitational force. Similar
treatments of this classical issue can be found elsewhere
\citep[e.g.][]{KB06}. However, given its great utility in establishing
the orbital evolution of galaxy mergers from a given set of initial
conditions, we have decided to include it in an Appendix in order to
provide a more complete view of our experiments.

Under the Keplerian approximation, the motion of a pair of gravitationally
bound galactic haloes, can be solved analytically by transforming it
into an equivalent one-body problem of a fictitious particle of mass
equal to the reduced mass of the system, defined by
\begin{equation}\label{reducedm}
\mu=\frac{M_{\rm p}M_{\rm s}}{M_{\rm p}+M_{\rm s}}\;,
\end{equation}
with $M_{\rm p}$ and $M_{\rm s}$ the virial masses of the host and
satellite galactic haloes, respectively, moving under the influence of
a central force equal to the gravitational interaction exerted by one
onto the other, and with a position vector ${\bf r}={\bf r}_{\rm
  p}-{\bf r}_{\rm s}$ whose modulus at $t=0$, $r_0$, gives the initial
relative distance between the point masses chosen to represent the
haloes (see. Along this work, quantities referred to
the more massive of the two merging galaxies (primary, host or
central) will be identified by the subindex 'p', while subindex 's',
will identify those referring to the less massive member of the pair
(secondary or satellite). Of course, in equal-mass mergers the
assignation of indexes is irrelevant.

Under these conditions, all the essential information about the initial motion of the system is contained on the orbit equation of the reduced body -- little by little, however, extended interacting galaxies will begin to deviate from this idealized orbit\footnote{Such divergence arises from two reasons: on one hand, the mutual gravitational acceleration of two spatially extended mass distributions is less than that of two equivalent point masses; on the other, dynamical friction induces orbital decay once the outer halo regions start to interpenetrate.} --, whose geometry is entirely specified by two constants of the motion, the total orbital energy
\begin{equation}\label{energy} 
{\mathcal E}=\frac{1}{2}\mu v^2-\frac{GM_{\rm p}M_{\rm s}}{r}\;, 
\end{equation} 
and angular momentum, whose magnitude is
\begin{equation} 
{\mathcal J}=\,\parallel \mu{\bf r}\times {\bf v}\parallel\,= \mu r v_{\rm tan}\;,
\end{equation} 
where we are using the notation
\begin{equation} 
{\bf v}=v_{\rm rad}\hat{\bf r}+v_{\rm tan}\hat{\pmb\theta}=\frac{dr}{dt}\hat{\bf r}+r\frac{d\theta}{dt}\hat{\pmb\theta}\;.
\end{equation} 
Equation~(\ref{energy}) can then be solved for the first derivative $dr/dt$ 
\begin{equation}\label{drdt} 
\frac{dr}{dt}=\sqrt{\frac{2}{\mu}}\left({\mathcal E}-\frac{1}{2}\frac{\mathcal J^2}{\mu r^2}+\frac{GM_{\rm p}M_{\rm s}}{r}\right)^{1/2}\;, 
\end{equation}
which can be integrated directly for $r(t)$. Since there are no external forces, the orbit of the reduced body lies in a plane with the angular momentum a constant vector pointing perpendicular to this plane. We note, however, that the complete solution of the equation of motion requires to provide, appart from the values of ${\mathcal E}$ and ${\mathcal J}$, one extra boundary condition at $t_0$. Since it is a very good idea to take $t_0=0$, the natural choices are therefore the modulus $r_0$ of the initial relative separation of the galaxies or of its time detivative $v_0$.

Alternatively, the two constants of the motion can be replaced by two
geometrical parameters, one describing the initial orbital
eccentricity
\begin{equation}\label{eccentricity}
e=\sqrt{1+\frac{2{\mathcal E}{\mathcal J}^2}{\mu(GM_{\rm p}M_{\rm s})^2}}\;,
\end{equation}
and other, the initial pericentric distance
\begin{equation}\label{perid}
r_{\rm per}(1+e)=\frac{{\mathcal J}^2}{\mu GM_{\rm p}M_{\rm s}}\;.
\end{equation}
For closed two-body systems (${\mathcal E}<0$) it is also possible to
replace the eccentricity and pericentric distance by two other fully
independent quantities: the circularity of the elliptical orbit, expressing the ratio between the lengths of its semi-major, $a$, and semi-minor, $b$, axes,
$\epsilon\equiv b/a =\sqrt{1-e^2}=j/j_{\rm cir}({\mathcal E})$, which is also a measure of 
the specific angular momentum of the orbit, $j\equiv rv_{\rm tan}$, relative to the
specific angular momentum of a circular orbit with the same energy,
and a parameter representing the initial orbital energy in
dimensionless form. In merger studies, it is customary to use for the
latter the ratio $r_{\rm cir,p}\equiv r_{\rm cir}({\mathcal E})/R_{\rm
  p}$, with $R_{\rm p}$ the virial radius of the primary halo, and
$r_{\rm cir}({\mathcal E})$ the radius of a circular orbit with the
same orbital energy ${\mathcal E}$. From the virial theorem:
\begin{equation}\label{circular}
r_{\rm cir}=-\frac{GM_{\rm p}M_{\rm s}}{2{\mathcal E}}\;.
\end{equation}
Finally, from equations~(\ref{eccentricity}) and (\ref{perid}), one gets
\begin{equation}
\frac{1-e}{r_{\rm per}}=-\frac{2{\mathcal E}}{GM_{\rm p}M_{\rm s}}\;,
\end{equation}
which combined with the definition (\ref{circular}) of the circular radius leads to
\begin{equation}\label{rperrcir}
r_{\rm per}\equiv a(1-e)=r_{\rm cir}(1-e)\;,
\end{equation}
a result that could have been anticipated because for circular orbits
the semi-major axis $a$ is equal to $r_{\rm cir}$. Given that $r_{\rm
  cir}$ is ${\mathcal O}(R_{\rm p})$ and the modal values of $e$ are
close to 1 (see below), this last relationship allows one to
understand why small pericentric separations are more frequent than
larger ones. Besides, since not
only the PDF of $e$ but also that of $r_{\rm per}$ are independent of
the mass ratio of the progenitors, equation~(\ref{rperrcir}) also
implies that the initial orbital energy as measured by $r_{\rm cir,p}$
must be equally universal for a given cosmological model. This is
consistent with the self-similarity of the galaxy formation process. 

Another aspect to take into consideration in the initial set-up of idealized binary merger simulations is the fact that it is a fairly widespread procedure to put the galaxies on parabolic orbits with a small pericentric distance (typically a small fraction of $R_{\rm p}$) that lead to fast merging. As noted by \citet{KB06}, this attempt to save computational time is, however, inconsistent with the outcome of cosmological simulations, which show
that most of the mergers are on orbits with ${\mathcal E}\lesssim 0$,
$e\lesssim 1$ and $r_{\rm per}\gtrsim 0.1R_{\rm p}$, independently of
the mass ratio of the progenitors. The fact that, to first order,
$r_{\rm per}\propto {\mathcal J}^2$ (equation~[\ref{perid}]) means
that by underestimating $r_{\rm per}$ one also underestimates the
amount of orbital angular momentum transferred to the remnant. This
may have a noticeable impact on its structure.  

Finally, it is important to keep in mind that the values of the
initial parameters involved in the Keplerian two-body problem cannot
be chosen in a completely arbitrary manner since certain combinations
of them do not lead to physically plausible results. Thus, for the
major mergers of the present study, the initial circularities and
separations adopted (see Section~\ref{ini_cond} and
Table~\ref{orb_params}) imply that $r_{\rm cir,p}\gtrsim 1$, as shown
in Fig.~\ref{fig_B1}.

%%%%%%%%%%%%%%%%%%%%%%%%%%%%%%%%%%%%%%%%%%%%%%%%%%

% Don't change these last lines
\label{lastpage}
\end{document}